\documentclass[12pt]{article}
\def\llsymbol#1{\@llsymbol{\@nameuse{c@#1}}}
\def\@llsymbol#1{\ifcase#1\or {}\or {'}\or {''}\or {'''}\or
   {''''}\or {'''''}\or  \else\@ctrerr\fi\relaz}
\newcounter{contador}
\newcommand{\letra}{
   \stepcounter{equation}
   \setcounter{contador}{\value{equation}}
   \setcounter{equation}{0}
   \renewcommand{\theequation}{\thecontador\alph{equation}}}
\newcommand{\antiletra}{
   \renewcommand{\theequation}{\arabic{equation}}
   \setcounter{equation}{\value{contador}}}

\begin{document}
\begin{center}{\Large \bf Ince's limits for confluent and
double-confluent Heun equations}
\vskip 0.4cm {B. D. Bonorino Figueiredo} \\
 Instituto de Cosmologia, Relatividade e Astrof\'isica (ICRA-BR)\\
{Centro Brasileiro de Pesquisas F\'{\i}sicas (CBPF)\\
 Rua Dr. Xavier Sigaud, 150 - 22290-180 - Rio de Janeiro, RJ, Brasil}
\end{center}
%
%
\begin{abstract}
\noindent
We find pairs of solutions to a differential equation
which is obtained as a special limit of a generalized
spheroidal wave equation (this is also known as confluent
Heun equation). One solution in each pair is
given by a series of hypergeometric
functions and converges for any finite value of the independent
variable $z$, while the other is given by
a series of modified Bessel
functions and converges for
$|z|>|z_{0}|$, where $z_{0}$ denotes a regular
singularity. For short, the preceding limit is called Ince's limit after
Ince who have used the same procedure to get the
Mathieu equations from the Whittaker-Hill ones.
We find as well that, when $z_{0}$ tends to zero, the
Ince limit of the generalized spheroidal wave equation turns
out to be the Ince limit of
a double-confluent Heun equation, for which solutions
are provided. Finally, we show that the Schr\"odinger equation
for inverse fourth and sixth-power potentials
reduces to peculiar cases of the double-confluent Heun
equation and its Ince's limit, respectively.
\end{abstract}
%
%
\section*{1. Introduction}
Firstly, we construct two linear differential equations
whose solutions behave at infinity as the so-called
subnormal Thom\'e solutions, in contrast to
solutions of a confluent and a
double-confluent Heun equations \cite{ronveaux}, from which
the former equations are obtained by a limit process. Secondly,
we provide solutions which afford the
expected asymptotic behavior for these equations.
Finally, we find that the Schr\"odinger equation
with inverse fourth and sixth-power potentials reduces to
particular instances of the double-confluent
Heun equation and its Ince limit, respectively.

In the first place, let us introduce the two equations under consideration.
Our starting point is the generalized spheroidal wave equation
(GSWE) in the form used by Leaver \cite{leaver1}, namely,
\begin{eqnarray}
\label{gswe1}
z(z-z_{0})\frac{d^{2}U}{dz^{2}}+(B_{1}+B_{2}z)
\frac{dU}{dz}+
\left[B_{3}-2\eta \omega(z-z_{0})+\omega^{2}z(z-z_{0})\right]U=0,
(\omega\neq 0)
\end{eqnarray}
where $B_{i}$, $\eta$ and $\omega$ are constants
(notice that, if $\omega=0$ and $\eta$ is fixed, the equation may
be transformed into a hypergeometric equation). The points
$z=0$ and $z=z_{0}$ are regular singularities
with indices ($0,1+B_{1}/z_{0}$)
and ($0,1-B_{2}-B_{1}/z_{0}$), respectively,
while the infinity is  an irregular singularity in which the
behavior of $U(z)$, inferred from the normal Thom\'e solutions
\cite{olver}, is given by
\begin{eqnarray}\label{asym}
\lim_{z\rightarrow  \infty}U(z)\sim e^{\pm i\omega z}z^{\mp i\eta-(B_{2}/2)}.
\end{eqnarray}
Since its parameters are not specified,
the above GSWE is equivalent to the
confluent Heun equation \cite{ronveaux}, an
equation that is more general than the original Wilson GSWE
\cite{wilson}. Furthermore, as noted by Leaver,
for $z_{0}=0$ we obtain a double-confluent Heun equation
(DCHE) having five parameters,
rather than  four as in other contexts \cite{decarreau1,schmidt},
namely,
\begin{eqnarray}\label{dche}
z^{2}\frac{d^{2}U}{dz^{2}}+(B_{1}+B_{2}z)\frac{dU}{dz}
+\left[B_{3}-2\eta \omega z+\omega^{2}z^{2}\right]U=0,\
(\omega\neq 0,\ B_{1}\neq 0),
\end{eqnarray}
where the singular points $z=0$ and $z=\infty$ are both irregular.
For $\omega= 0$ and/or $B_{1}= 0$ (with $\eta$ fixed) this equation
degenerates into confluent hypergeometric equations (see Appendix A).
At infinity, the behavior of $U(z)$ is again given by
(\ref{asym}), while at $z=0$ we find in the usual way
\cite{olver} that
\begin{eqnarray}\label{z0}
\lim_{z\rightarrow  0}U(z)\sim 1 \ \ \mbox{or}\ \
\lim_{z\rightarrow  0}U(z)\sim e^{B_{1}/ z}z^{2-B_{2}} .
\end{eqnarray}
The Leaver procedure also allows us to obtain solutions to
the DCHE from solutions to the GSWE when $z_{0}$ goes
to zero. The known Leaver-type solutions \cite{leaver1,eu}
are appropriate to solve,
for instance, the Teukolsky equations for the extreme upper limit of
the rotation parameter \cite{leaver1}, the time dependence
of Dirac test fields in dust-dominated
Friedmann-Robertson-Walker spacetimes
and the Schr\"odinger equation with asymmetric double-Morse
potentials \cite{eu}. They are suitable either
for handling the Schr\"odinger equation with inverse fourth-power
potentials, as we will see.

Now, to get the equations we are interested in,
the Levear limit is combined with a limit
that Ince \cite{ince} had
used to derive the Mathieu equation from the Whittaker-Hill
equation. The Ince limit is obtained by taking
\begin{eqnarray}\label{limits}
\omega\rightarrow 0, \ \
\eta\rightarrow
\infty, \ \mbox{such that }\  \ 2\eta \omega =-q,
\end{eqnarray}
where $q$ is a constant. Thus, the Ince limit of the GSWE is
\begin{eqnarray}
\label{lindemann}
z(z-z_{0})\frac{d^{2}U}{dz^{2}}+(B_{1}+B_{2}z)
\frac{dU}{dz}+
\left[B_{3}+q(z-z_{0})\right]U=0,\ (q\neq0).
\end{eqnarray}
This is a generalization of the Mathieu equation for, by setting
\letra
\begin{eqnarray}\label{mathieu2}
z_{0}=1,\ B_{1}=-1/2, \ B_{2}=1,\ z=\cos^2{(\sigma u)},\
W(u)=U(z),
\end{eqnarray}
we obtain the equation
\begin{eqnarray}\label{mathieu}
\frac{d^2W}{du^2}+\sigma^2\left[2q-4B_{3}-2q
\cos(2\sigma u)\right]W=0,
\end{eqnarray}
that is, the Mathieu equation if $\sigma=1$,
and the modified Mathieu equation if $\sigma=i$
\cite{McLachlan}. In fact,  inserting $ z_{0}=1$,
$\ B_{1}=-1/2$ and $\ B_{2}=1 $ into Eq. (\ref{lindemann}),
one recovers the algebraic Lindemann form
for the Mathieu equation \cite{lindemann}. Nevertheless, the trigonometric
form (\ref{mathieu})  with  $4B_{3}=2q-a$ is useful to
verify that our solutions for the Ince limit of the GSWE  give
solutions already known for the Mathieu equation.

On the other hand, the Ince limit of the DCHE -- or Leaver limit of
the Eq. (\ref{lindemann}) -- is the equation
\antiletra
\begin{eqnarray}
\label{lindemann2}
z^2\frac{d^{2}U}{dz^{2}}+(B_{1}+B_{2}z)
\frac{dU}{dz}+
\left(B_{3}+qz\right)U=0,\ (q\neq0, \ B_{1}\neq 0)
\end{eqnarray}
which degenerates into simpler equations if $q=0$
and/or $B_{1}=0$ (see Appendix A). Solutions are obtained
for this special DCHE by taking the Leaver limit
of solutions for the Eq. (\ref{lindemann}). By the way, we shall
see that the Schr\"odinger equation for
an inverse sixth-power potential is a particular case
of Eq. (\ref{lindemann2}), as stated in the first paragraph.

We emphasize that the Ince limits of
the GSWE and DCHE, unlike the original GSWE and DCHE,
require solutions behaving according to
the subnormal Thom\'e solutions \cite{olver}, that is,
\begin{eqnarray}\label{thome2}
\lim_{z\rightarrow  \infty}U(z)\sim
e^{\pm 2i\sqrt{qz}}z^{(1/4)-(B_{2}/2)}.
\end{eqnarray}
Despite this, our main mathematical issue consists in deriving
pairs of series solutions to Eqs.  (\ref{lindemann}) and (\ref{lindemann2})
-- having the behavior stipulated above at the singular points --
from pairs of solutions to the GSWE. For this we shall
again employ the Ince and Leaver limits. The solutions in each pair
have the same series coefficients and these satisfy
three-term recurrence relations.

In section 2, a pair of solutions for the Ince limit of the GSWE
is obtained by taking the Ince limit (\ref{limits})
of a known pair of solutions  for the GSWE. One solution is
given by an expansion in series of hypegeometric functions and
converges for any finite $z$; the other solution is
given by an expansion in series of modified
Bessel functions and converges for
$|z|>|z_{0}|$. Other
pairs are generated by using transformation
rules. These rules result from variable
substitutions that preserve the form of
the differential equations but
modify their parameters and/or arguments.

In section 3,  we find pairs of solutions for the Ince
limit  of the DCHE  by taking the
Leaver limit ($z_{0}\rightarrow0$) of solutions for the
Ince limit of the GSWE. Solutions in series
of irregular confluent hypergeometric functions result
from expansions in series of hypergeometric functions
and converge for any finite $z$. The other solution in each pair
is given by a series of modified Bessel functions and
converges for $\vert {z}\vert>0$.

In both of these sections we deal with solutions with and
without a phase parameter $\nu$. In general, this $\nu$ is introduced
in order to assure the convergence of the series when
there is no free constant in the differential
equation, as in some scattering problems
or in equations where $z$ is a variable related to
the time \cite{leaver1,eu}.
Solutions with a phase parameter are
two-sided in the sense that the summation index $n$
runs from $-\infty$ to $\infty$. However, if there is an arbitrary
parameter in the equation, we can
truncate the series by requiring that
$n\geq 0$. In this manner, we obtain
$\nu$ in terms of parameters of
the differential equation.

In section 4, we show that the Schr\"odinger
equation with inverse fourth and sixth-power
potentials in fact leads to the DCHE and its Ince limit.
Some additional considerations are provided in section 5,
while in Appendix A we discuss the degenerate cases of the
DCHEs, in Appendix B we present an alternative derivation
of the expansions
in Bessel functions, and in Appendix C we rewrite the Leaver-type
solutions for the DCHE in a form appropriate to solve
the Schr\"odinger equation with an inverse fourth-power potential.

%
%
\section*{2. Ince's limits for the generalized spheroidal wave equation }
In this section we use transformation rules that permit us to generate
new solutions from a given solution for the Ince limit of the GSWE.
The rules $T_{1}$, $T_{2}$ and $T_{3}$ below
are derived from the ones valid for the GSWE
 \cite{eu} and can be checked by
substitution of variables.
If $U(z)=U(B_{1},B_{2},B_{3};
z_{0},q;z)$ denotes one solution for Eq.
(\ref{lindemann}), the effects of
these rules are as follows
\letra
\begin{eqnarray}\begin{array}{l}
T_{1}U(z)=z^{1+B_{1}/z_{0}}
U(C_{1},C_{2},C_{3};z_{0},q;z),\  \ z_{0}\neq0,
\vspace{3mm}\\
T_{2}U(z)=(z-z_{0})^{1-B_{2}-B_{1}/z_{0}}U(B_{1},D_{2},D_{3};
z_{0},q;z), \ \  z_{0}\neq0,
\vspace{3mm}\\
T_{3}U(z)=
U(-B_{1}-B_{2}z_{0},B_{2},
B_{3}-q z_{0};z_{0},-q;z_{0}-z),
\end{array}
\end{eqnarray}
where
\begin{eqnarray}\begin{array}{l}
C_{1}=-B_{1}-2z_{0}, \ \
C_{2}=2+B_{2}+\frac{2B_{1}}{z_{0}},\ C_{3}=B_{3}+
\left(1+\frac{B_{1}}{z_{0}}\right)
\left(B_{2}+\frac{B_{1}}{z_{0}}\right),
\vspace{.3cm}\\
D_{2}=2-B_{2}-\frac{2B_{1}}{z_{0}},\ D_{3}=B_{3}+
\frac{B_{1}}{z_{0}}\left(\frac{B_{1}}{z_{0}}
+B_{2}-1\right).
\end{array}
\end{eqnarray}
We use only $T_{1}$ and
$T_{2}$. The rule $T_{3}$ exchange
the position of the regular singular points $z=z_{0}
\leftrightarrow z=0$ and may be used to get
an alternative representation for the solutions,
but these are not proper
for getting the limit $z_{0}\rightarrow 0$.

In section 2.1 we derive two pairs of solutions  for the Ince
limit of the GSWE -- denoted by $ (U_{i\nu}^{0},
U_{i\nu}^{\infty})$,  $i=1,2$ -- with a phase parameter $\nu$.
The superscript `zero' indicates
that the series converges in any finite part of the complex
plane, while the superscript `infinity' indicates convergence
for $\vert z\vert>\vert z_{0}\vert$.
The second pair of solutions results from the first by
means of the rule $T_{2}$.
In section 2.2, we truncate these series by taking $n\geq0$
and obtain four pairs of solutions without
phase parameter.
%
%
%
\subsection*{2.1. Solutions with a phase parameter}
Denoting by $b_{n}$ the series coefficients of a solution, their
recurrence relations will have the general form
\antiletra\letra
\begin{eqnarray}\label{rec}
\alpha_{n}b_{n+1}+\beta_{n}b_{n}+\gamma_{n}b_{n-1}=0,\ (-\infty<n<\infty)
\end{eqnarray}
where $\alpha_{n}$, $\beta_{n}$, $\gamma_{n}$
and $b_{n}$ depend on a phase parameter $\nu$
which may be determined from a characteristic
equation given as a sum of two infinite continued fractions, namely,
\begin{eqnarray}
\beta_{0}=\frac{\alpha_{-1}\gamma_{0}}{\beta_{-1}-} \frac{\alpha_{-2}
\gamma_{-1}}{\beta_{-2}-}\frac{\alpha_{-3}\gamma_{-2}}
{\beta_{-3}-}\cdots+\frac{\alpha_{0}\gamma_{1}}{\beta_{1}-}
\frac{\alpha_{1}\gamma_{2}}
{\beta_{2}-}\frac{\alpha_{2}\gamma_{3}}{\beta_{3}-}\cdots .
\end{eqnarray}
For a specific pair of solutions we add
a superscript in each of these quantities.

The first pair of solutions for the Ince limit of the GSWE comes from
the following pair of solutions of the GSWE \cite{eu}
\antiletra\letra
\begin{eqnarray}
\label{eu} \begin{array}{l}
U_{1\nu}^{0}(z)=
e^{i\omega z}\displaystyle \sum_{n=-\infty}^{\infty}b_{n}^{(1)}F\left(\frac{B_{2}}{2}-n-\nu-1,
n+\nu+\frac{B_{2}}{2};B_{2}+\frac{B_{1}}{z_{0}};
1-\frac{z}{z_{0}}\right),
\vspace{3mm}\\
U_{1\nu}^{\infty}(z) =e^{i\omega z}z^{1-(B_{2}/2)}
\displaystyle \sum_{n=-\infty}^{\infty}b_{n}^{(1)}
(-2i\omega z)^{n+\nu}
\Psi(n+\nu+1+i\eta,2n+2\nu+2;-2i\omega z),
\end{array}\end{eqnarray}
where $F(a,b;c;y)$ and $\Psi(a,b;y)$ denote, respectively,
the hypergeometric functions and the irregular confluent
hypergeometric functions \cite{abramowitz,erdelyi1}. The solution
$U_{1\nu}^{0}$ converges for any finite $z$, whereas
$U_{1\nu}^{\infty}$ converges for $\mid z\mid>\mid z_{0}\mid$.
In the recurrence relations (\ref{rec}) for $b_{n}^{(1)}$ we have
\begin{eqnarray}
\begin{array}{l}
\alpha_{n}^{(1)}  =  i\omega z_{0}\frac{\left(n+\nu+2-\frac{B_{2}}{2}\right)
\left(n+\nu+1-\frac{B_{2}}{2}-\frac{B_{1}}{z_{0}}\right)
\left(n+\nu+1-i\eta\right)}
{2(n+\nu+1)\left(n+\nu+\frac{3}{2}\right)},
\vspace{3mm} \\
\beta_{n}^{(1)}  =  -B_{3}-\eta \omega z_{0}-(n+\nu+1-\frac{B_{2}}{2})
(n+\nu+\frac{B_{2}}{2})
-\frac{\eta \omega z_{0}\left(\frac{B_{2}}{2}-1\right)
\left(\frac{B_{2}}{2}+\frac{B_{1}}{z_{0}}\right)}
{(n+\nu)(n+\nu+1)},
\vspace{.3cm} \\
\gamma_{n}^{(1)}  = -i\omega z_{0}\frac{\left(n+\nu+\frac{B_{2}}{2}-1\right)
\left(n+\nu+\frac{B_{2}}{2}+\frac{B_{1}}{z_{0}}\right)
(n+\nu+i\eta)}
{2\left(n+\nu-\frac{1}{2}\right)(n+\nu)}.
\end{array}
\end{eqnarray}
Note that $\nu$
cannot be integer or half-integer in order
to avoid vanishing denominators in
the coefficients of the recurrence relations.
Moreover, for an integer or half-integer $\nu$,
we would have two equal hypergeometric or
confluent hypergeometric
functions (for different values of $n$),
contrary to the hypothesis that all the terms
of the series are independent. On the other
hand, the hypergeometric functions
are not defined if $B_{2}+(B_{1}/z_{0})$
is zero or a negative integer. Nonetheless,
a transformation rule supplies another solution
which is valid for these values of $B_{2}+(B_{1}/z_{0})$.

The three-term recurrence relations (\ref{rec}) constitute
a infinite system of homogeneous linear equations for which
nontrivial solutions for the coefficients $b_{n}$ demand
that the determinant of respective tridiagonal matrix vanishes.
Equivalently, the characteristic equation must be satisfied and
this is a condition necessary also to assure the convergence of the series
by means of a Poincar\'e-Perron theorem \cite{gautschi}.
However, there are two possibilities to satisfy this requirement.

On the one hand, if there is some free constant in the differential
equation, that constant must be determined so that the characteristic
equation is fulfilled for the admissible values of $\nu$ (that is,
neither integer nor half-integer). In this case, the freedom of
choosing $\nu$ may be used in two different ways: (i) to
obtain two-sided solutions ($-\infty<n<\infty$) by ascribing
appropriate values for $\nu$, or (ii) to obtain one-sided
solutions by choosing $\nu$ such that $n\geq 0$. At the end
of the present section, we use the first
alternative to rederive some Poole's solutions \cite{poole,poole2} for
the Mathieu equation, having period $2\pi m$, where $m$ is any integer
equal or greater than 2. In section 2.2, we use the second
alternative for the general case.
These latter solutions afford solutions with period $\pi$ or $2\pi$
for the Mathieu equation, in contrast to the solutions obtained
in the first alternative.

On the other hand, if there is no arbitrary parameter in the
differential equation, the parameter $\nu$ takes the
role of free parameter in the sense that it must be
adjusted to ensure the validity of the characteristic equation
and, consequently, the convergence of the series. By this
reason, $\nu$ is also called characteristic index or parameter
\cite{buhring2}. Examples of equations requiring a phase
parameter are discussed in section 4.

From the above pair of solutions for the GSWE,
by using the Ince limit (\ref{limits}), we readily find the solution
$U_{1\nu}^{0}(z)$ written in the first pair below.
To get the Ince limit of the solution $U_{1\nu}^{\infty}(z)$,
we define $c_{n}$ as
\begin{eqnarray*}
b_{n}^{(1)}=(i\eta)^{n+\nu}\Gamma(i\eta-n-\nu)c_{n}.
\end{eqnarray*}
This imply that
\begin{eqnarray*}
U_{1\nu}^{\infty}(z) =e^{i\omega z}z^{1-(B_{2}/2)}
\displaystyle \sum_{n=-\infty}^{\infty}c_{n}\Gamma(i\eta-n-\nu)
(-qz)^{n+\nu}
\Psi\left(n+\nu+1+i\eta,2n+2\nu+2;-\frac{qz}{i\eta}\right),
\end{eqnarray*}
where $q=-2\eta \omega$. The recurrence relations for $c_{n}$
are
\begin{eqnarray*}
\bar{\alpha}_{n}c_{n+1}+\beta_{n}^{(1)}c_{n}+\bar{\gamma}_{n}c_{n-1}=0,\ (-\infty<n<\infty)
\end{eqnarray*}
with
\antiletra
\begin{eqnarray*}
\bar{\alpha}_{n}=\frac{i\eta}{i\eta-n-\nu-1}\alpha_{n}^{(1)}, \
\bar{\gamma}_{n}=\frac{i\eta-\nu-1}{i\eta}\gamma_{n}^{(1)}.
\end{eqnarray*}
On the other hand, we have \cite{erdelyi1}
\begin{eqnarray}
\lim_{a\rightarrow \label{K} \infty}[\Gamma(a+1-c)\Psi(a,c;x/a)]=
2x^{(1-c)/2}K_{c-1}(2\sqrt{x})
\end{eqnarray}
where $K_{\lambda}(\xi)$ denotes the
modified Bessel function of the second
kind \cite{luke} whose definition in terms of
irregular confluent hypergeometric functions
is \cite{erdelyi1}
\begin{eqnarray}\label{2.8}
K_{\lambda}(\xi)=K_{-\lambda}(\xi)=\sqrt{\pi}\ e^{-\xi}(2\xi)^{\lambda}
\Psi\left(\lambda+\frac{1}{2},2\lambda+1;2\xi\right).
\end{eqnarray}
Then, using (\ref{K}) we find that for $i\eta\rightarrow \infty$
($n$ fixed and $q=$constant)
\begin{eqnarray*}\begin{array}{l}
\Gamma(i\eta-n-\nu)
(-qz)^{n+\nu}
\Psi\left(n+\nu+1+i\eta,2n+2\nu+2;-\frac{qz}{i\eta}\right)
\rightarrow 2(-qz)^{1/2}K_{2n+2\nu+1}(\pm 2i\sqrt{qz}),
\vspace{3mm}\\
\lim\bar{\alpha}_{n}\rightarrow \lim\alpha_{n}^{(1)},\ \
\lim\bar{\gamma}_{n}\rightarrow \lim\gamma_{n}^{(1)}\Rightarrow
\lim c_{n}\rightarrow \lim b_{n}^{(1)}.
\end{array}\end{eqnarray*}
Using these results, we find the Ince limit of $U_{1\nu}^{\infty}$,
written in the first pair below. Although this is a formal derivation, the
solution may be checked by inserting it into Eq. (\ref{lindemann})
(see Appendix B). In addition, from the relation \cite{luke}
\begin{eqnarray}\label{21}
\lim_{\mid \xi\mid\rightarrow  \infty}K_{\lambda}(\xi)\sim
\sqrt{\frac{\pi}{2\xi}}\ e^{-\xi}, \ \  -\frac{3\pi}{2}<\arg\xi<
\frac{3\pi}{2}
\end{eqnarray}
we see that the expansions in series of
Bessel functions have the behavior given by
\begin{eqnarray*}
\lim_{z\rightarrow \infty}U_{j\nu}^{\infty}(z)\sim
e^{\pm 2i\sqrt{qz}}z^{(1/4)-(B_{2}/2)}, \ \  -\frac{3\pi}{2}
<\arg(\pm2i\sqrt{qz})<
\frac{3\pi}{2},\ \ (j=1,2)
\end{eqnarray*}
in accordance with Eq. (\ref{thome2}).
The second pair of solutions  follows
from the first one through the rule $T_{2}$, as mentioned before.
Moreover, solutions for the Mathieu
equation are obtained
by using Eqs. (\ref{mathieu2}) and by
noting that in this case the hypergeometric
functions can be rewritten in terms of trigonometric functions.\\

%

\noindent
{\bf First pair}
\letra
\begin{eqnarray}
\label{s1}
\begin{array}{l}
U_{1\nu}^{0}(z)=
\displaystyle \sum_{n=-\infty}^{\infty}b_{n}^{(1)}F\left(\frac{B_{2}}{2}-n-\nu-1,
n+\nu+\frac{B_{2}}{2};B_{2}+\frac{B_{1}}{z_{0}};
1-\frac{z}{z_{0}}\right),
\vspace{.3cm}\\
U_{1\nu}^{\infty}(z) =z^{(1-B_{2})/2}
\displaystyle \sum_{n=-\infty}^{\infty} b_{n}^{(1)}
K_{2n+2\nu+1}\left(\pm2i\sqrt{qz}\right),
\end{array}
\end{eqnarray}
where in the recurrence relations (\ref{rec})
\begin{eqnarray}\label{apB}
\begin{array}{l}
\alpha_{n}^{(1)}  =  q z_{0}\frac{\left(n+\nu+2-\frac{B_{2}}{2}\right)
\left(n+\nu+1-\frac{B_{2}}{2}-\frac{B_{1}}{z_{0}}\right)}
{(n+\nu+1)\left(n+\nu+\frac{3}{2}\right)},
\vspace{.2cm} \\
\beta_{n}^{(1)}  =  4B_{3}-2q z_{0}+4\left(n+\nu+1-\frac{B_{2}}{2}\right)
\left(n+\nu+\frac{B_{2}}{2}\right)
-2q z_{0}\frac{\left(\frac{B_{2}}{2}-1\right)
\left(\frac{B_{2}}{2}+\frac{B_{1}}{z_{0}}\right)}
{(n+\nu)(n+\nu+1)},
\vspace{.3cm} \\
\gamma_{n}^{(1)} = q z_{0}\frac{\left(n+\nu+\frac{B_{2}}{2}-1\right)
\left(n+\nu+\frac{B_{2}}{2}+\frac{B_{1}}{z_{0}}\right)}
{\left(n+\nu-\frac{1}{2}\right)(n+\nu)}.
\end{array}
\end{eqnarray}
If $B_{2}+(B_{1}/z_{0})$ is zero or a negative
integer we have the solution $U_{2\nu}^{0}$
instead of $U_{1\nu}^{0}$.

For the Mathieu equation we use Eqs. (\ref{mathieu2}) and
the formula \cite{abramowitz}
\antiletra
\begin{eqnarray*}
\label{hyper1}
F\left[-a,a; (1/2);\sin^2(\sigma u)\right]=\cos(2a\sigma u).
\end{eqnarray*}
Thence, we obtain even solutions with respect to
$u$, namely,
\letra
\begin{eqnarray}
\begin{array}{ll}
W_{1\nu}^{0}(u)=
\displaystyle \sum_{n=-\infty}^{\infty}b_{n}^{(1)}\cos[(2n+2\nu+1)\sigma u],& \
|\cos(\sigma u)|< \infty,
\vspace{.3cm}\\
W_{1\nu}^{\infty}(u) =
\displaystyle \sum_{n=-\infty}^{\infty} b_{n}^{(1)}
K_{2n+2\nu+1}\left[\pm2i\sqrt{q}\cos(\sigma u)\right],&
|\cos(\sigma u)|> 1,
\end{array}
\end{eqnarray}
with the simplified recurrence relations ($a=2q-4B_{3}$)
\begin{eqnarray}
qb_{n+1}^{(1)}+\left[\left(
2n+2\nu+1\right)^2-a\right]b_{n}^{(1)}+qb_{n-1}^{(1)}=0.
\end{eqnarray}
%

\noindent
{\bf Second pair}
\antiletra
\letra
\begin{eqnarray}
\begin{array}{l}
U_{2\nu}^{0}(z)=(z-z_{0})^{1-B_{2}
-\frac{B_{1}}{z_{0}}}\ z^{1+\frac{B_{1}}{z_{0}}}
\displaystyle \sum_{n=-\infty}^{\infty}b_{n}^{(2)}\times
\vspace{.3cm}\\
\hspace{1.5cm}F\left(-n-\nu-\frac{B_{2}}{2}+1,
n+\nu+2-\frac{B_{2}}{2};
2-B_{2}-\frac{B_{1}}{z_{0}};1-\frac{z}{z_{0}}\right),
\vspace{.3cm}\\
U_{2\nu}^{\infty}(z)=(z-z_{0})^{1-B_{2}-\frac{B_{1}}{z_{0}}}
\ z^{\frac{B_{1}}{z_{0}}+\frac{B_{2}}{2}-\frac{1}{2}}
\displaystyle \sum_{n=-\infty}^{\infty}b_{n}^{(2)}
K_{2n+2\nu+1}\left(\pm2i\sqrt{qz}\right),
\end{array}
\end{eqnarray}
where
\begin{eqnarray}
\begin{array}{l}
\alpha_{n}^{(2)}=  qz_{0}\frac{\left(n+\nu+\frac{B_{2}}{2}\right)
\left(n+\nu+1+\frac{B_{2}}{2}+
\frac{B_{1}}{z_{0}}\right)}{(n+\nu+1)
\left(n+\nu+\frac{3}{2}\right)},
\  \beta_{n}^{(2)}= \beta_{n}^{(1)},
\vspace{.3cm} \\
\gamma_{n}^{(2)}= q z_{0}\frac{\left(n+\nu+1-\frac{B_{2}}{2}\right)
\left(n+\nu-\frac{B_{2}}{2}-
\frac{B_{1}}{z_{0}}\right)}
{\left(n+\nu-\frac{1}{2}\right)(n+\nu)},
\end{array}
\end{eqnarray}
in the recurrence relations (\ref{rec}) for $b_{n}^{(2)}$.
If $B_{2}+(B_{1}/z_{0})$ is a positive integer
equal or greater than $2$ we have the
solution $U_{1\nu}^{0}$ instead of
$U_{2\nu}^{0}$. Note that, in writing the solution
$U_{2\nu}^{0}$, we have used the relation
\antiletra
\begin{eqnarray}\label{hyp2}
F(a,b;c;y)=(1-y)^{c-a-b}F(c-a,c-b;c;y).
\end{eqnarray}

For the Mathieu equation we use the relation
\cite{abramowitz}
\begin{eqnarray*}
\label{hyper2}
F\left(a,1-a;\frac{3}{2};
\sin^2(\sigma u)\right)=\frac{\sin[(2a-1)\sigma u]}
{(2a-1)\sin(\sigma u)}
\end{eqnarray*}
and, in addition, define
$c_{n}$ as $b_{n}^{(2)}=(2n+2\nu+1)c_{n}$.
So, we find that the recurrence relations
for $c_{n}$ become identical to the ones
for $b_{n}^{(1)}$, giving the odd solutions
\begin{eqnarray}
\begin{array}{ll}
W_{2\nu}^{0}(u)=
\displaystyle \sum_{n=-\infty}^{\infty}b_{n}^{(1)}\sin[(2n+2\nu+1)\sigma u],
\vspace{.3cm}\\
W_{2\nu}^{\infty}(u) =\tan{(\sigma u)}
\displaystyle \sum_{n=-\infty}^{\infty}\left(2n+2\nu+1
\right) b_{n}^{(1)}
K_{2n+2\nu+1}\left[\pm2i\sqrt{q}\cos(\sigma u)\right],
\end{array}
\end{eqnarray}
where $|\cos(\sigma u)|< \infty$ and
$|\cos(\sigma u)|> 1$, respectively.

As we have explained earlier, if there is a free parameter
in the differential equation, it is possible to satisfy the
characteristic equation for any noninteger or half-integer $\nu$.
We use this fact to rederive some Poole's solutions \cite{poole,poole2}
to the Mathieu equation. For this, in the previous $W_{1\nu}^{0}(u)$
 and  $W_{1\nu}^{0}(u)$ we take
 \begin{eqnarray}
2\nu+1=l/m, \ \ \sigma=1,
\end{eqnarray}
where $l$ and $m$ are integers prime to one another, $l<m$.
Then, we find the two-sided Poole solutions $W_{1}^{P}(u)$
 and  $W_{1}^{P}(u)$ given by
 \begin{eqnarray}
W_{1}^{P}(u)=
\displaystyle \sum_{n=-\infty}^{\infty}b_{n}^{(1)}\cos
\left[ \left( 2n+\frac{l}{m}\right)  u\right] ,\ \
W_{2}^{P}(u)=
\displaystyle \sum_{n=-\infty}^{\infty}b_{n}^{(1)}\sin
\left[ \left( 2n+\frac{l}{m}\right)  u\right] .
\end{eqnarray}
The first is even with respect to $u$ and the second is odd,
and both of them have period $2\pi m$, $m>1$. Since they have
the same series coefficients, we can combine them to
find another Poole solution, that is,
 \begin{eqnarray}
W^{P}(u)=
\displaystyle \sum_{n=-\infty}^{\infty}b_{n}^{(1)}\exp
\left[ i\left( 2n+\frac{l}{m}\right)  u\right] .
\end{eqnarray}
Furthermore, for an arbitrary $\nu$ we find
 \begin{eqnarray}
W(u)=
\displaystyle \sum_{n=-\infty}^{\infty}b_{n}^{(1)}\exp
\left[ i\left( 2n+2\nu+1\right)  u\right] ,
\end{eqnarray}
which is also a solution already known in the literature
\cite{abramowitz, poole2}.
%
%
%
%
%
%
%
\subsection*{2.2. Solutions without phase parameter}
Now we truncate the solutions obtained in section 2.1
by taking $n\geq 0$. This gives $\nu$ in terms of some parameters
of the differential equation. The resulting solutions are
convergent only if there is a free parameter to be determined
from the characteristic equation.

This truncation reverses the procedure by which the
solution $U_{1\nu}^{0}(z)$ for the GSWE, given
in Eq. (\ref{eu}), was obtained. Indeed, that solution
was constructed  \cite{eu2} as a generalization of an one-sided series
of Jacobi polynomials, constructed by Fackerell and Crossman
\cite{fackerell} to solve the angular Teukolsky equations of
the relativistic astrophysics.
Despite this, the truncated solutions found in \cite{eu} are more general
than the Fackerell-Crossman ones because no particular
values are attached to the parameters of the GSWE
and also because the truncation was extended to the Leaver
expansion $U_{1\nu}^{\infty}(z)$. I addition, these one-sided
series are suitable to get solutions in finite series,
the so called quasi-polynomial solutions. In effect, a solution
whose coefficients $b_{n}$ obey recurrence relations as
\begin{eqnarray*}
\alpha_{n}b_{n+1}+\beta_{n}b_{n}+\gamma_{n}b_{n-1}=0,
\ \ n\geq0,\ \ b_{-1}=0
\end{eqnarray*}
becomes a quasi-polynomial solution with $0\leq n\leq N-1$ whenever
$\gamma_{N}=0$ for some $n=N$ \cite{arscott2} .

For the truncated solutions -- denoted by
($U_{i}^{0},U_{i}^{\infty}$), $i=1,2,3,4$ --
the recurrence relations and the
characteristic equations have one of
the three forms written below. The first
case ($\alpha_{-1}=0$) is the general
one and the others ($\alpha_{-1}\neq 0$)
may occur only for special cases.
\begin{eqnarray}
&&\left.
\begin{array}{l}
\alpha_{0}b_{1}+\beta_{0}b_{0}=0,
\vspace{.2cm} \\
\alpha_{n}b_{n+1}+\beta_{n}b_{n}+
\gamma_{n}b_{n-1}=0\ (n\geq1),
\end{array}\right\} \Rightarrow
\beta_{0}=\frac{\alpha_{0}\gamma_{1}}{\beta_{1}-}\ \frac{\alpha_{1}
\gamma_{2}}
{\beta_{2}-}\ \frac{\alpha_{2}\gamma_{3}}{\beta_{3}-}\cdots.
\label{r1a}
\end{eqnarray}
\begin{eqnarray}
\left.
\begin{array}{l}
\alpha_{0}b_{1}+\beta_{0}b_{0}=0,
\vspace{.2cm} \\
\alpha_{1}b_{2}+\beta_{1}b_{1}+\left[
\alpha_{-1}+\gamma_{1}\right]b_{0}=0,
\vspace{.2cm} \\
\alpha_{n}b_{n+1}+\beta_{n}b_{n}+\gamma_{n}
b_{n-1}=0\ (n\geq2),
\end{array}\right\}\Rightarrow
\beta_{0}=\frac{\alpha_{0}\left[\alpha_{-1}+\gamma_{1}
\right]}
{\beta_{1}-}
\ \frac{\alpha_{1}\gamma_{2}}{\beta_{2}-}\ \frac{\alpha_{2}\gamma_{3}}
{\beta_{3}-}\cdots .
\label{r2a}
\end{eqnarray}
\begin{eqnarray}
\left.
\begin{array}{l}
\alpha_{0}b_{1}+\left[\beta_{0}+\alpha_{-1}
\right]b_{0}=0,
\vspace{.2cm} \\
\alpha_{n}b_{n+1}+\beta_{n}b_{n}
+\gamma_{n}b_{n-1}=0\ (n\geq1),
\end{array}\right\}\Rightarrow
\beta_{0}+\alpha_{-1}=\frac{\alpha_{0}\gamma_{1}}
{\beta_{1}-}\ \frac{\alpha_{1}\gamma_{2}}{\beta_{2}-}
\ \frac{\alpha_{2}\gamma_{3}}
{\beta_{3}-}\cdots .
\label{r3a}
\end{eqnarray}
Note that we have
$n\geq -1$ in $\alpha_{n}$, $n\geq0$ in $\beta_{n}$ and
$n\geq 1$ in $\gamma_{n}$.

These forms for the recurrence relations are the same
that appear in truncation of the expansions (\ref{eu})
for the GSWE \cite{eu}. As a matter of fact,
the solutions of the present section are the Ince limit of
solutions for  the GSWE given in section 3 of Ref. \cite{eu}.
However, in order to illustrate how these recurrence relations are
obtained, we insert the solution  $U_{1\nu}^{\infty}$ given in (\ref{s1})
into the Ince limit of the GSWE. Then, for $n\geq 0$, from
Eq. (\ref{2.10}) we find that
\begin{eqnarray*}
\displaystyle \sum_{n=0}^{\infty}
\alpha_{n-1}^{(1)}b_{n}^{(1)}
K_{2n+2\nu-1}(\xi)+
\displaystyle \sum_{n=0}^{\infty}
\beta_{n}^{(1)}b_{n}^{(1)}
K_{2n+2\nu+1}(\xi)+
\displaystyle \sum_{n=0}^{\infty}
\gamma_{n+1}^{(1)}b_{n}^{(1)}
K_{2n+2\nu+3}(\xi)=0.
\end{eqnarray*}
Setting  $m=n-1$, $m=n$ and $m=n+1$ in the first, second and
third terms, respectively, this equation becomes
\begin{eqnarray}\label{recorrencia}
&&\alpha_{-1}b_{0}K_{2\nu-1}(\xi)+
\left[\alpha_{0}b_{1}+\beta_{0}
b_{0}\right]K_{2\nu+1}(\xi)+
\left[\alpha_{1}b_{2}+\beta_{1}
b_{1}+\gamma_{1}b_{0}\right]K_{2\nu+3}(\xi)
+\nonumber\\
&&
\displaystyle \sum_{n=2}^{\infty}\left[\alpha_{n}b_{n+1}
+\beta_{n}b_{n}+
\gamma_{n}b_{n-1}\right]K_{2n+2\nu+1}(\xi)=0,
\end{eqnarray}
where we have dropped the upper suffixes.
Therefore, if we can choose $\nu$ so
that $\alpha_{-1}=0$, we find the first set
of recurrence relations. However, notice that
\begin{eqnarray*}
\alpha_{-1}=\frac{qz_{0} \left(\nu+1-\frac{B_{2}}{2}\right)
\left(\nu-\frac{B_{1}}{x_{0}}-\frac{B_{2}}{2}\right)
}{2\nu(\nu+1/2)}=0,\ \mbox{if}
\left\{
\begin{array}{l}
\nu=\frac{B_{2}}{2}-1
\ \mbox{for}\ B_{2}\neq 1,2;
\vspace{3mm}\\
\nu=\frac{B_{1}}{x_{0}}+\frac{B_{2}}{2} \
\ \mbox{for} \    \frac{B_{1}}{x_{0}}+\frac{B_{2}}{2}\neq0,
\frac{1}{2}.
\end{array}
\right.
\end{eqnarray*}
Hence we see that there are two possible choices
for $\nu$ and, for each of them we have two
cases in which $\alpha_{-1}$ may differ from
zero. Let us consider only the case $\nu=(B_{2}/2)-1$.
Then, for the exceptional case $B_{2}=1$ ($\nu=-1/2$),
we find $K_{2\nu-1}=K_{2\nu+3}=K_{2}$ (since
$K_{\lambda}=K_{-\lambda}$) and therefore the Bessel
functions in the first and third terms of Eq.
(\ref{recorrencia}) are equal, giving the
recurrence relations
(\ref{r2a}). Similarly, if $B_{2}=2$ ($\nu=0$)
we find $K_{2\nu-1}=K_{2\nu+1}=K_{1}$ and
this leads to the recurrence relations (\ref{r3a}).
In this manner we obtain the first pair given
below. The remaining can be derived from this
by using the transformations rules $T_{1}$ and $T_{2}$ as
\begin{eqnarray*}
\left(U_{1}^{0},U_{1}^{\infty}\right)
\stackrel{T_{1}}{\longleftrightarrow}
\left(U_{2}^{0},U_{2}^{\infty}\right)
\stackrel{T_{2}}{\longleftrightarrow}
\left(U_{3}^{0},U_{3}^{\infty}\right)
\stackrel{T_{1}}{\longleftrightarrow}
\left(U_{4}^{0},U_{4}^{\infty}\right)
\stackrel{T_{2}}{\longleftrightarrow}
\left(U_{1}^{0},U_{1}^{\infty}\right).
\end{eqnarray*}
The condition on each pair is imposed
in order to assure that the special functions
are independent in both solutions;
it guarantees either that there is no
vanishing denominator in the recurrence
relations. Furthermore, we have additional restrictions
on the parameters of the solutions
$U_{i}^{0}$. Thus, if $B_{2}+(B_{1}/z_{0})$
is zero or a negative integer, the
hypergeometric functions are not defined in
$U_{1}^{0}$ and $U_{2}^{0}$ but are defined
in $U_{3}^{0}$ and $U_{4}^{0}$, and vice-versa.
The results for the Mathieu
equations are already known \cite{McLachlan},
but note that the recurrence relations for this case come
from the three Eqs. (\ref{r1a}-\ref{r3a}) above.\\
%
%

\noindent
{\bf First pair}: $B_{2}\neq 0,-1,-2,\cdots$. This first pair
corresponds to $\nu=(B_{2}/2)-1$ in
($U_{1\nu}^{ 0},U_{1\nu}^{\infty }$).
\letra
\begin{eqnarray}
\begin{array}{l}
U_{1}^{0}(z)=
\displaystyle \sum_{n=0}^{\infty}b_{n}^{(1)}F\left(-n,
n+B_{2}-1;B_{2}+\frac{B_{1}}{z_{0}};
1-\frac{z}{z_{0}}\right),
\vspace{.3cm}\\
U_{1}^{\infty}(z) =z^{(1-B_{2})/2}
\displaystyle \sum_{n=0}^{\infty} b_{n}^{(1)}
K_{2n+B_{2}-1}\left(\pm2i\sqrt{qz}\right),
\end{array}
\end{eqnarray}
with the following coefficients
\begin{eqnarray}
\begin{array}{l}
\alpha_{n}^{(1)}  =  \frac{q z_{0}\left(n+1\right)
\left(n-\frac{B_{1}}{z_{0}}\right)}
{\left(n+\frac{B_{2}}{2}\right)\left(n+\frac{B_{2}}{2}+\frac{1}{2}
\right)},
\vspace{.2cm} \\
\beta_{n}^{(1)}  =  4B_{3}-2q z_{0}+4n\left(n+B_{2}-1\right)
-\frac{2q z_{0}\left(\frac{B_{2}}{2}-1\right)
\left(\frac{B_{2}}{2}+\frac{B_{1}}{z_{0}}\right)}
{\left(n+\frac{B_{2}}{2}-1\right)
\left(n+\frac{B_{2}}{2}\right)},
\vspace{.3cm} \\
\gamma_{n}^{(1)} = \frac{q z_{0}\left(n+B_{2}-2\right)
\left(n+B_{2}+\frac{B_{1}}{z_{0}}-1\right)}
{\left(n+\frac{B_{2}}{2}-\frac{3}{2}\right)
\left(n+\frac{B_{2}}{2}-1\right)},
\end{array}
\end{eqnarray}
in the recurrence relations for the
$b_{n}^{(1)}$, namely: Eqs. (\ref{r1a}) if
$B_{2}\neq 1,2$; Eqs. (\ref{r2a}) if $B_{2}=1$;
Eqs. (\ref{r3a}) if $B_{2}=2$.

For the Mathieu equation we find solutions
\antiletra
\letra
\begin{eqnarray}
\begin{array}{ll}
W_{1}^{0}(u)=
\displaystyle \sum_{n=0}^{\infty}b_{n}^{(1)}\cos(2n\sigma u),& |\cos(\sigma u)|< \infty,
\vspace{.3cm}\\
W_{1}^{\infty}(u) =
\displaystyle \sum_{n=0}^{\infty} b_{n}^{(1)}
K_{2n}\left[\pm2i\sqrt{q}\cos(\sigma u)\right],&
|\cos(\sigma u)|> 1,
\end{array}
\end{eqnarray}
with the simplified recurrence relations
\begin{eqnarray}\begin{array}{l}
qb_{1}^{(1)}-ab_{0}^{(1)}=0,
\ \  qb_{2}^{(1)}+\left[4-a\right]b_{1}^{(1)}+2qb_{0}^{(1)}=0,\vspace{.3cm}\\
qb_{n+1}^{(1)}+\left[
4n^2-a\right]b_{n}^{(1)}+qb_{n-1}^{(1)}=0\ (n\geq2).
\end{array}
\end{eqnarray}
These solutions are even with respect to $u$ and, for $\sigma=1$,
the solution $W_{1}^{0}(u)$ has period $\pi$.\\

%
%
\noindent
{\bf Second pair}: $(B_{2}/2)+(B_{1}/z_{0})\neq-1,
-3/2,-2,-5/2\cdots$. This pair of solutions can also be obtained by
taking $\nu=(B_{2}/2)+(B_{1}/z_{0})$ in
($U_{1\nu}^{ 0},U_{1\nu}^{0 }$).
\antiletra
\letra
\begin{eqnarray}
\begin{array}{l}
U_{2}^{0}(z)=z^{1+(B_{1}/z_{0})}
\displaystyle \sum_{n=0}^{\infty}b_{n}^{ (2)}
F\left(-n,n+1+B_{2}+\frac{2B_{1}}{z_{0}};B_{2}+
\frac{B_{1}}{z_{0}};1-\frac{z}{z_{0}}\right),
\vspace{.3cm}\\
U_{2}^{\infty}(z)=z^{(1-B_{2})/2}
\displaystyle \sum_{n=0}^{\infty}b_{n}^{(2)}
K_{2n+1+B_{2}+(2B_{1}/z_{0})}\left(\pm2i\sqrt{qz}
\right),
\end{array}
\end{eqnarray}
where
\begin{eqnarray}
\begin{array}{l}
\alpha_{n}^{^{(2)}}=  \frac{qz_{0} (n+1)
\left(n+2+\frac{B_{1}}{x_{0}}\right)}
{\left(n+1+\frac{B_{2}}{2}+\frac{B_{1}}{z_{0}}\right)
\left(n+\frac{3}{2}+\frac{B_{2}}{2}+\frac{B_{1}}{z_{0}}
\right)},
\vspace{.3cm} \\
\beta_{n}^{^{(2)}}= 4B_{3}-2q z_{0}+4\left(n+1+\frac{B_{1}}{x_{0}}\right)
\left(n+B_{2}+\frac{B_{1}}{x_{0}}\right)
-\frac{2q z_{0}\left(\frac{B_{2}}{2}-1\right)
\left(\frac{B_{2}}{2}+\frac{B_{1}}{z_{0}}\right)}
{\left(n+\frac{B_{2}}{2}+\frac{B_{1}}{x_{0}}\right)
\left(n+1+\frac{B_{2}}{2}+\frac{B_{1}}{x_{0}}\right)},
\vspace{.2cm} \\
\gamma_{n}^{^{(2)}}= \frac{q z_{0}\left(n+B_{2}+
\frac{B_{1}}{x_{0}}-1\right)
\left(n+B_{2}+\frac{2B_{1}}{z_{0}}\right)}
{\left(n-\frac{1}{2}+\frac{B_{2}}{2}+\frac{B_{1}}
{z_{0}}\right)
\left(n+\frac{B_{2}}{2}+\frac{B_{1}}{z_{0}}\right)},
\end{array}
\end{eqnarray}
in the recurrence relations for $b_{n}^{(2)}$:
Eqs. (\ref{r1a})
if $(B_{2}/2)+(B_{1}/z_{0})\neq 0,
-1/2$; Eqs. (\ref{r2a}) if
$(B_{2}/2)+(B_{1}/z_{0})=
-1/2$; Eqs. (\ref{r3a}) if
$(B_{2}/2)+(B_{1}/z_{0})=0$.

For the Mathieu equation we again have even solutions
\antiletra
\letra
\begin{eqnarray}
\begin{array}{ll}
W_{2}^{0}(u)=
\displaystyle \sum_{n=0}^{\infty}b_{n}^{(2)}\cos[(2n+1)\sigma u],& |\cos(\sigma u)|< \infty,
\vspace{.3cm}\\
W_{2}^{\infty}(u) =
\displaystyle \sum_{n=0}^{\infty} b_{n}^{(2)}
K_{2n+1}\left[\pm2i\sqrt{q}\cos(\sigma u)\right],&
|\cos(\sigma u)|> 1,
\end{array}
\end{eqnarray}
with the recurrence relations
\begin{eqnarray}\begin{array}{l}
qb_{1}^{(2)}+\left[q+1-a\right]b_{0}^{(2)}=0, \vspace{.3cm}\\
qb_{n+1}^{(2)}+\left[\left(
2n+1\right)^2-a\right]b_{n}^{(2)}+qb_{n-1}^{(2)}=0 \ (n\geq 1).
\end{array}
\end{eqnarray}
If $\sigma=1$ the solution $W_{2}^{0}(u)$
has period $2\pi$.\\

%
%
%
%
\noindent
{\bf Third pair}: $B_{2}\neq 4,5,6,\cdots$.
This corresponds to $\nu=1-(B_{2}/2)$ in ($U_{2\nu}^{0}, U_{2\nu}^{\infty})$.
\antiletra
\letra
\begin{eqnarray}
\begin{array}{l}
U_{3}^{0}(z)=
(z-z_{0})^{1-B_{2}-\frac{B_{1}}{z_{0}}}
z^{1+\frac{B_{1}}{z_{0}}}
\displaystyle \sum_{n=0}^{\infty}b_{n}^{(3)}
F\left(-n,n+3-B_{2};2-B_{2}-\frac{B_{1}}{z_{0}};
1-\frac{z}{z_{0}}\right),
\vspace{3mm}\\
U_{3}^{\infty}(z) =(z-z_{0})^{1-B_{2}-
\frac{B_{1}}{z_{0}}}z^{\frac{B_{1}}{z_{0}}+\frac{B_{2}}{2}-\frac{1}{2}}
\displaystyle \sum_{n=0}^{\infty}b_{n}^{(3)}
K_{2n+3-B_{2}}\left(\pm2i\sqrt{qz}\right),
\end{array}
\end{eqnarray}
with the coefficients
\begin{eqnarray}
\begin{array}{l}
\alpha_{n}^{ (3)} =  \frac{qz_{0}\
(n+1)
\left(n+2+\frac{B_{1}}{z_{0}}\right)}
{\left(n+2-\frac{B_{2}}{2}\right)\left(n+\frac{5}{2}-\frac{B_{2}}{2}\right)},
\vspace{.2cm} \\
\beta_{n}^{ (3)}  =  4B_{3}-2q z_{0}+4(n+1)(n+2-B_{2})
-\frac{2q z_{0}\left(\frac{B_{2}}{2}-1\right)
\left(\frac{B_{2}}{2}+\frac{B_{1}}{z_{0}}\right)}
{\left(n+1-\frac{B_{2}}{2}\right)\left(n+2-
\frac{B_{2}}{2}\right)},
\vspace{.3cm} \\
\gamma_{n}^{ (3)}  =
\frac{q z_{0}\ \left(n+2-B_{2}\right)
\left(n+1-B_{2}-\frac{B_{1}}{z_{0}}\right)}
{\left(n+\frac{1}{2}-\frac{B_{2}}{2}\right)\left(n+1-\frac{B_{2}}{2}\right)}.
\end{array}
\end{eqnarray}
in the recurrence relations: Eqs. (\ref{r1a}) if
$B_{2}\neq 2,3$; Eqs. (\ref{r2a}) if $B_{2}=3$;
Eqs. (\ref{r3a}) if $B_{2}=2$.

For the Mathieu equation we redefine the coefficients
$b_{n}^{(3)}$ as $b_{n}^{(3)}\rightarrow (2n+2)b_{n}^{(3)}$. Then
we find the odd solutions
\antiletra
\letra
\begin{eqnarray}
\begin{array}{ll}
W_{3}^{0}(u)=
\displaystyle \sum_{n=0}^{\infty}b_{n}^{(3)}\sin[(2n+2)\sigma u],& |\cos(\sigma u)|< \infty,
\vspace{.3cm}\\
W_{3}^{\infty}(u) =\tan{(\sigma u)}
\displaystyle \sum_{n=0}^{\infty}\left(2n+2
\right) b_{n}^{(3)}
K_{2n+2}\left[\pm2i\sqrt{q}\cos(\sigma u)\right],&
|\cos(\sigma u)|> 1,
\end{array}
\end{eqnarray}
with the recurrence relations
\begin{eqnarray}\begin{array}{l}
qb_{1}^{(3)}+\left[4-a\right]b_{0}^{(3)}=0,
\vspace{.3cm}\\
qb_{n+1}^{(3)}+\left[4\left(
n+1\right)^2-a\right]b_{n}^{(3)}+qb_{n-1}^{(3)}=0,
\ (n\geq 1).
\end{array}
\end{eqnarray}
For $\sigma=1$ the solution $W_{3}^{0}(u)$
has period $\pi$.\\

%
%
%
%
\noindent
{\bf Fourth pair}:  $(B_{2}/2)+(B_{1}/z_{0})\neq1,
3/2,2,5/2\cdots$. This can also be obtained by
setting $\nu=-(B_{2}/2)-(B_{1}/z_{0})$ in
($U_{2\nu}^{ 0},U_{2\nu}^{ \infty}$)
\antiletra
\letra
\begin{eqnarray}
\begin{array}{l}
U_{4}^{0}=(z-z_{0})^{1-B_{2}-\frac{B_{1}}{z_{0}}}
\displaystyle \sum_{n=0}^{\infty}b_{n}^{ (4)}
F\left(-n,n+1-B_{2}-\frac{2B_{1}}{z_{0}};2-B_{2}
-\frac{B_{1}}{z_{0}};1-\frac{z}{z_{0}}\right), \vspace{5mm}\\
U_{4}^{\infty} =(z-z_{0})^{1-B_{2}-\frac{B_{1}}{z_{0}}}
z^{\frac{B_{1}}{z_{0}}+\frac{B_{2}}{2}-\frac{1}{2}}\displaystyle \sum_{n=0}^{\infty}b_{n}^{ (4)}
K_{2n+1-B_{2}-(2B_{1}/z_{0})}\left(\pm2i\sqrt{qz}\right),
\end{array}
\end{eqnarray}
with coefficients
\begin{eqnarray}
\begin{array}{l}
\alpha_{n}^{ (4)}  =  \frac{qz_{0}\ (n+1)
\left(n-\frac{B_{1}}{z_{0}}\right)}
{\left(n+1-\frac{B_{2}}{2}-\frac{B_{1}}{z_{0}}\right)
\left(n+\frac{3}{2}-\frac{B_{2}}{2}-\frac{B_{1}}{z_{0}}
\right)},
\vspace{.2cm} \\
\beta_{n}^{ (4)}  =  4B_{3}-2q z_{0}+4
\left(n-\frac{B_{1}}{z_{0}}\right)
\left(n-B_{2}+1-\frac{B_{1}}{z_{0}}\right)
-\frac{2q z_{0}\left(\frac{B_{2}}{2}-1\right)
\left(\frac{B_{2}}{2}+\frac{B_{1}}{z_{0}}\right)}
{\left(n-\frac{B_{2}}{2}-\frac{B_{1}}{z_{0}}\right)
\left(n+1-\frac{B_{2}}{2}-\frac{B_{1}}{z_{0}}\right)},
\vspace{.3cm} \\
\gamma_{n}^{ (4)}  = \frac{q z_{0}\
\left(n+1-B_{2}-\frac{B_{1}}{z_{0}}\right)
\left(n-B_{2}-\frac{2B_{1}}{z_{0}}\right)}
{\left(n-\frac{1}{2}-\frac{B_{2}}{2}-\frac{B_{1}}
{z_{0}}\right)
\left(n-\frac{B_{2}}{2}-\frac{B_{1}}{z_{0}}\right)},
\end{array}
\end{eqnarray}
in the recurrence relations:  Eqs. (\ref{r1a}) if
$(B_{2}/2)+(B_{1}/z_{0})\neq 0, 1/2$;
Eqs. (\ref{r2a}) if $(B_{2}/2)+(B_{1}/z_{0})=1/2$;
Eqs. (\ref{r3a}) if $(B_{2}/2)+(B_{1}/z_{0})=0$.

For the Mathieu equation we redefine
$b_{n}(4)$ according to $b_{n}^{(4)}\rightarrow (2n+1)b_{n}^{(4)}$
and find the odd solutions
\antiletra
\letra
\begin{eqnarray}
\begin{array}{ll}
W_{4}^{0}(u)=
\displaystyle \sum_{n=0}^{\infty}b_{n}^{(4)}\sin[(2n+1)\sigma u],& |\cos(\sigma u)|< \infty,
\vspace{.3cm}\\
W_{4}^{\infty}(u) =\tan{(\sigma u)}
\displaystyle \sum_{n=0}^{\infty}\left(2n+1
\right) b_{n}^{(4)}
K_{2n+1}\left[\pm2i\sqrt{q}\cos(\sigma u)\right],&
|\cos(\sigma u)|> 1,
\end{array}
\end{eqnarray}
with the recurrence relations
\begin{eqnarray}\begin{array}{l}
qb_{4}^{(4)}+\left[1-q-a\right]b_{0}^{(4)}=0,
\vspace{.3cm}\\
qb_{n+1}^{(4)}+\left[\left(
2n+1\right)^2-a\right]b_{n}^{(4)}+qb_{n-1}^{(4)}=0 \ (n\geq1).
\end{array}
\end{eqnarray}
Now, for $\sigma=1$, $W_{4}^{0}(u)$ has period $2\pi$.
%
%
%
\section*{3. Ince's limits for the double-confluent Heun equation}
As in the case of the Ince limit of the GSWE, we have found no solution
in the literature for the Ince limit of the DCHE.
The solutions below are obtained by taking the limit
$z_{0}\rightarrow 0$ (Leaver limit) of the solutions
given in section 2 for the Ince limit of the GSWE.
For this we use the formulas \cite{erdelyi1}
\antiletra\letra
\begin{eqnarray}\label{rel}
&&\lim_{c\rightarrow \infty}F\left(a,b;c;1-\frac{c}{y}\right)=
y^a\Psi(a,a+1-b;y),\\
&&\lim_{\alpha\rightarrow \infty}\left(1+\frac{y}{\alpha}\right)^{\alpha}=e^{y}
\Rightarrow \lim_{z_{0}\rightarrow 0}\left(1-\frac{z_{0}}{z}\right)^{-B_{1}/
z_{0}}=e^{B_{1}/z}.
\end{eqnarray}
Actually, it is not necessary to use the second equation above,
since we can get one pair of solutions as the limit of the first
pair of section 2.1
and, then, generate the other pair by means of the transformation rule
\antiletra
\begin{eqnarray}
\tau U(z)=e^{{B_{1}}/{z}}z^{2-B_{2}}U(-B_{1},4-B_{2},
B_{3}+2-B_{2}; q;z),
\end{eqnarray}
where $U(z)=U(B_{1},B_{2},B_{3}; q;z)$ denotes known
solutions of Eq.  (\ref{lindemann2}).
On the other hand, to check that the solutions $U_{i}^{0}(z)$ exhibit
the behavior given in Eq. (\ref{z0}) when $z\rightarrow 0$,
we may use the relation \cite{erdelyi1}
\begin{eqnarray}\label{tricomi}
\lim_{\vert y\vert\rightarrow \infty}\Psi(a,b;y)\sim y^{-a}[1+O(|y|^{-1}],
\ \ -\frac{3\pi}{2}<\arg y< \frac{3\pi}{2}.
\end{eqnarray}
%
%
%
\subsection*{3.1. Solutions with a phase parameter}
For the solution $U_{1\nu}^{0}$ of
section 2.1, we find that the limit
of the hypergeometric functions when $z_{0}$
tends to zero, $B_{2}$ and $B_{1}$ being fixed ($c=B_{2}+B_{1}/z_{0}
\rightarrow \infty$),  is given by
\begin{eqnarray*}
&&\lim_{z_{0}\rightarrow 0}F\left(n+\nu+\frac{B_{2}}{2},-n-\nu-1+
\frac{B_{2}}{2};B_{2}+\frac{B_{1}}{z_{0}};
1-\frac{z}{z_{0}}\right)\\
&&\propto z^{-\nu-\frac{B_{2}}{2}}
\left(\frac{B_{1}}{z}\right)^{n}\Psi\left(n+\nu+\frac{B_{2}}{2},2n+2\nu+2;
\frac{B_{1}}{z}\right).
\end{eqnarray*}
Then, considering also the solution $U_{1\nu}^{\infty}$ and the
limits for the coefficients in the recurrence relations, we get the first pair
of solutions with a phase parameter $\nu$ (different
of integer or half-integer). The rule $\tau$ leads
to the second pair.\\

%
%
\noindent
{\bf First pair}
\letra
\begin{eqnarray}
\begin{array}{l}
U_{1\nu}^{0}(z)=
z^{-\nu-\frac{B_{2}}{2}}\displaystyle \sum_{n=-\infty}^{\infty}
b_{n}^{(1)}
\left(\frac{B_{1}}{z}\right)^{n}\Psi\left(n+\nu+
\frac{B_{2}}{2},2n+2\nu+2;
\frac{B_{1}}{z}\right),
\vspace{.3cm}\\
U_{1\nu}^{\infty}(z) =z^{(1-B_{2})/2}
\displaystyle \sum_{n=-\infty}^{\infty} b_{n}^{(1)}
K_{2n+2\nu+1}\left(\pm2i\sqrt{qz}\right),
\end{array}
\end{eqnarray}
where in the recurrence relations (\ref{rec})
\begin{eqnarray}
\begin{array}{l}
\alpha_{n}^{(1)}  =  - \frac{qB_{1}\left(n+\nu+2-\frac{B_{2}}{2}\right)}
{(n+\nu+1)\left(n+\nu+\frac{3}{2}\right)},
\vspace{.2cm} \\
\beta_{n}^{(1)}  =  4B_{3}+4\left(n+\nu+1-\frac{B_{2}}{2}\right)
\left(n+\nu+\frac{B_{2}}{2}\right)
-\frac{q B_{1}\left(B_{2}-2\right)}
{(n+\nu)(n+\nu+1)},
\vspace{.3cm} \\
\gamma_{n}^{(1)} =  \frac{qB_{1}\left(n+\nu+\frac{B_{2}}{2}-1\right)}
{\left(n+\nu-\frac{1}{2}\right)(n+\nu)}.
\end{array}
\end{eqnarray}

%
%
\noindent
{\bf Second pair}
\antiletra
\letra
\begin{eqnarray}
\begin{array}{l}
U_{2\nu}^{0}(z)=e^{{B_{1}}/{z}}z^{-\nu-
\frac{B_{2}}{2}}\displaystyle \sum_{n=-\infty}^{\infty}b_{n}^{(2)}
\left(-\frac{B_{1}}{z}\right)^{n}\Psi\left(n+\nu+2-
\frac{B_{2}}{2},2n+2\nu+2;
-\frac{B_{1}}{z}\right),
\vspace{.3cm}\\
U_{2\nu}^{\infty}(z)=e^{{B_{1}}/{z}}z^{(1-B_{2})/2}
\displaystyle \sum_{n=-\infty}^{\infty}b_{n}^{(2)}
K_{2n+2\nu+1}\left(\pm2i\sqrt{qz}\right),
\end{array}
\end{eqnarray}
where
\begin{eqnarray}
\begin{array}{l}
\alpha_{n}^{^{(2)}}=  \frac{qB_{1}\left(n+\nu+\frac{B_{2}}{2}\right)}
{(n+\nu+1)\left(n+\nu+\frac{3}{2}\right)}, \
\beta_{n}^{^{(2)}}= \beta_{n}^{(1)},\
\gamma_{n}^{^{(2)}}= -\frac{qB_{1}\left(n+\nu+1-\frac{B_{2}}{2}\right)}
{\left(n+\nu-\frac{1}{2}\right)(n+\nu)},
\end{array}
\end{eqnarray}
in the recurrence relations (\ref{rec}) for $b_{n}^{(2)}$.

%
%
\subsection*{3.2. Solutions without phase parameter}
These solutions may be derived by truncating the solutions
of section 3.1. In this case, we see that there is only one choice for
$\nu$ in each pair. Alternatively, the solutions can be found
by applying the Leaver procedure to the first and third pairs of
section 2.2.\\

%
%

\noindent
{\bf First pair}: $B_{2}\neq 0,-1,-2,\cdots$.
This corresponds to $\nu=(B_{2}/2)-1$ in $\left(U_{1\nu}^{ 0},
U_{1\nu}^{\infty }\right)$.
\antiletra
\letra
\begin{eqnarray}
\begin{array}{l}
U_{1}^{0}(z)=z^{1-B_{2}}
\displaystyle \sum_{n=0}^{\infty}b_{n}^{(1)}\left(\frac{B_{1}}
{z}\right)^n\Psi\left(n+B_{2}-1,2n+B_{2};
\frac{B_{1}}{z}\right),
\vspace{.3cm}\\
U_{1}^{\infty}(z) =z^{(1-B_{2})/2}
\displaystyle \sum_{n=0}^{\infty} b_{n}^{(1)}
K_{2n+B_{2}-1}\left(\pm2i\sqrt{qz}\right),
\end{array}
\end{eqnarray}
with the following coefficients
\begin{eqnarray}
\begin{array}{l}
\alpha_{n}^{(1)}  = - \frac{q B_{1}\left(n+1\right)}
{\left(n+\frac{B_{2}}{2}\right)\left(n+\frac{B_{2}}{2}+\frac{1}{2}
\right)},
\vspace{.2cm} \\
\beta_{n}^{(1)}  =  4B_{3}+4n\left(n+B_{2}-1\right)
-\frac{q B_{1}\left(B_{2}-2\right)}
{\left(n+\frac{B_{2}}{2}-1\right)
\left(n+\frac{B_{2}}{2}\right)},
\vspace{.3cm} \\
\gamma_{n}^{(1)} = \frac{q B_{1}\left(n+B_{2}-2\right)}
{\left(n+\frac{B_{2}}{2}-\frac{3}{2}\right)
\left(n+\frac{B_{2}}{2}-1\right)}.
\end{array}
\end{eqnarray}
in the recurrence relations for the
$b_{n}^{(1)}$: Eqs. (\ref{r1a}) if
$B_{2}\neq 1,2$; Eqs. (\ref{r2a}) if
$B_{2}=1$; Eqs. (\ref{r3a}) if $B_{2}=2$.\\

%
%
%
\noindent
{\bf Second  pair}: $B_{2}\neq 4,5,6,\cdots$.
It corresponds to $\nu=1-(B_{2}/2)$ in $\left(U_{2\nu}^{ 0},
U_{2\nu}^{\infty }\right)$ but can also be obtained from the first
pair via the rule $\tau$.
\antiletra
\letra
\begin{eqnarray}
\begin{array}{l}
U_{2}^{0}(z)=e^{{B_{1}}/{z}}z^{-1}\displaystyle \sum_{n=0}^{\infty}b_{n}^{(2)}
\left(-\frac{B_{1}}{z}\right)^{n}\Psi\left(n+3-B_{2},2n+4-B_{2};
-\frac{B_{1}}{z}\right),
\vspace{.3cm}\\
U_{2}^{\infty}(z)=e^{{B_{1}}/{z}}z^{(1-B_{2})/2}\displaystyle \sum_{n=0}^{\infty}b_{n}^{(2)}
K_{2n+3-B_{2}}\left(\pm2i\sqrt{qz}\right),
\end{array}
\end{eqnarray}
where
\begin{eqnarray}
\begin{array}{l}
\alpha_{n}^{ (3)} =  \frac{qB_{1}\
(n+1)}
{\left(n+2-\frac{B_{2}}{2}\right)\left(n+\frac{5}{2}-\frac{B_{2}}{2}\right)},
\vspace{.2cm} \\
\beta_{n}^{ (3)}  =  4B_{3}+4(n+1)(n+2-B_{2})
-\frac{qB_{1}\left(B_{2}-2\right)}
{\left(n+1-\frac{B_{2}}{2}\right)\left(n+2-
\frac{B_{2}}{2}\right)},
\vspace{.3cm} \\
\gamma_{n}^{ (3)}  =
-\frac{q B_{1}\ \left(n+2-B_{2}\right)}
{\left(n+\frac{1}{2}-\frac{B_{2}}{2}\right)\left(n+1-\frac{B_{2}}{2}\right)}.
\end{array}
\end{eqnarray}
\antiletra
in the recurrence relations for $b_{n}^{(2)}$: Eqs. (\ref{r1a}) if
$B_{2}\neq 2,3$; Eqs. (\ref{r2a}) if $B_{2}=3$;
Eqs. (\ref{r3a}) if $B_{2}=2$.
%

%
\section*{4. Potential applications}
As we have mentioned, the Schr\"odinger equation
with inverse fourth and sixth-power potentials can be
reduced, respectively, to the double-confluent Heun
equation (\ref{dche}) and its Ince limit (\ref{lindemann2}).
Singular potentials like these have appeared in the description
of intermolecular forces \cite{frank} and in the
scattering of ions by polarizable atoms. For the sake
of illustration, we consider the last problem.

Before discussing these examples, let us present the so called
normal forms of the DCHE, that is, the forms in which there is no
first-order derivative terms in the differential equations. The general
procedure for this, consists in writing the equation as
\begin{eqnarray*}
\frac{d^2 U}{dz^2}+p(z)\frac{dU}{dz}+q(z)U=0.
\end{eqnarray*}
Then, the substitution
\begin{eqnarray*}
U(z)=F(z)\exp{\left(-\frac{1}{2}\int p(z)dz\right)}
\end{eqnarray*}
gives a first normal form, namely,
\begin{eqnarray*}
\frac{d^2 F}{dz^2}+I(z)F=0,\ \ I(z)=q(z)-\frac{1}{2}\frac{dp(z)}{dz}
-\frac{1}{4}[p(z)]^2.
\end{eqnarray*}
From this, other normal forms are obtained by the transformations
\begin{eqnarray*}
z=h(\vartheta),\ \ F(z)=\sqrt{\frac{dh}{d\vartheta}}\ G(\vartheta)
\end{eqnarray*}
which yield
\begin{eqnarray*}
\frac{d^2 G}{d\vartheta^2}+J(\vartheta)G=0,\ \ J(\vartheta)=I[h(\vartheta)]\left( \frac{dh}{d\vartheta}\right) ^2+
\frac{1}{2}\frac{d^3h}{d\vartheta^3}/
\frac{dh}{d\vartheta}-
\frac{3}{4}\left( \frac{d^2h}{d\vartheta^2}
/ \frac{dh}{d\vartheta}\right)^2  .
\end{eqnarray*}

By employing this procedure, Lemieux and Bose \cite{lemieux}
have derived several normal forms for the general Heun equation
and its confluent cases, excepting the triconfluent equation.
These forms are useful to recognize whether a given equation
belongs to the Heun class. Nevertheless, to find the solutions
for the equation, we have to come back to the form for which
 the solutions were established, as below.
The three Lemieux-Bose normal forms for the DCHE, together
 with the transformations of variables, are the following:
\begin{eqnarray}\label{N1}
\begin{array}{l}
U(z)=z^{-B_{2}/2}e^{B_{1}/(2z)}F(z),
\vspace{3mm}\\
\frac{d^2F}{dz^2}+\left[\omega^{2}-\frac{2\eta \omega}{z}+\frac{1}{z^2}
\left(B_{3}-\frac{B_{2}^{2}}{4}+\frac{B_{2}}{2}\right)+\frac{B_{1}}{z^3}
\left(1-\frac{B_{2}}{2}\right)
-\frac{B_{1}^2}{4z^4}\right]F=0;
\end{array}
\end{eqnarray}
\begin{eqnarray}\label{N2}
\begin{array}{l}
z= \rho^2, \ \ U(z)=\rho^{(1-2B_{2})/2}e^{B_{1}/(2\rho^2)}G(\rho)
\Leftrightarrow G(\rho)=z^{(2B_{2}-1)/4} e^{-B_{1}/(2z)}U(z),
\vspace{3mm}\\
\frac{d^2G}{d\rho^2}+\left[4\omega^{2}\rho^2-8\eta \omega+\frac{4}{\rho^2}
\left(B_{3}-\frac{B_{2}^{2}}{4}+\frac{B_{2}}{2}-\frac{3}{16}\right)+
\frac{4B_{1}}{\rho^4}
\left(1-\frac{B_{2}}{2}\right)
-\frac{B_{1}^2}{\rho^6}\right]G=0;
\end{array}
\end{eqnarray}
\begin{eqnarray}\label{N3}
\begin{array}{l}
z=e^{\lambda u},\ \
U(z)=H(u)\exp\left[ {\frac{1}{2}\lambda(1-B_{2}) u+\frac{B_{1}}{2}e^{-\lambda u}}\right] \ \Leftrightarrow
\ H(u)=z^{(B_{2}-1)/2}e^{-B_{1}/(2z)}U(z),
\vspace{3mm}\\
\frac{d^2H}{du^2}+
\lambda^2\left[ B_{3}-\left( \frac{1-B_{2}}{2}\right)^2
-\frac{B_{1}^2}{4}e^{-2\lambda u} -
B_{1}\left( \frac{B_{2}}{2}-1\right)e^{-\lambda u}-
2\eta \omega e^{\lambda u} +\omega^2e^{2\lambda u}\right] H=0,
\end{array}
\end{eqnarray}
where $\lambda$ is a constant at our disposal, for example,
$\lambda=1$ or $\lambda=i$. Note that, since these transformations
involve neither $\eta$ nor $\omega$, their Ince limits
are obtained by putting $\omega^2=0$ and $2\eta\omega=-q$.

Now we proceed with the scattering problem. The radial
part $R(r)=\chi(r)/r$ of the wave function for the
Schr\"{o}dinger equation in three dimensions,
for a particle with mass $\mu$ and energy $E$, is
\begin{eqnarray}\label{radial}
\frac{d^2\chi(r)}{dr^2}+\left[k^2-\frac{l(l+1)}{r^2}-
\frac{2\mu}{\hbar^{2}}V(r)
\right]\chi(r)=0,
\end{eqnarray}
where $k^2=2\mu E/\hbar^2$, $l$ is the angular momentum and $V(r)$
is the potential. Now, according to Kleinman,
Hahn and Spruch \cite{kleinman}, for the interaction of a light
particle  of charge $e'$ with a fixed atom of charge $Z\overline{e}$
containing $z'$ electrons, we have
\begin{eqnarray}
V(r)=\frac{(Z-z')\overline{e}e'}{r}-\frac{\alpha_{1}'{(e')^{2}}}{2r^4}-
\left(\alpha_{2}'-6a_{0}\beta_{1}'\right)
\frac{(e')^2}{2r^6},
\end{eqnarray}
where $r$ is the distance from the incident ion to the atom,
$a_{0}=\hbar^2/(\mu\overline{e}^{2} )$ is the Bohr radius,
$\alpha_{1}'$ and $\alpha_{2}'$ are,
respectively, the electric dipole and quadrupole polarizabilities
of the atom and  $\beta_{1}'$ is a  parameter
resulting from a nonadiabatic correction ($\alpha_{1}'$, $\alpha_{2}'$
and $\beta_{1}'$ are constants which describe the properties
of the target only).  For this potential,  the Schr\"odinger
equation becomes
\begin{eqnarray}\label{radial2}
\frac{d^2\chi}{dr^2}+\left[k^2-\frac{2\mu(Z-z')
\overline{e}e'}{\hbar^{2} r}
-\frac{l(l+1)}{ r^2}
+\frac{\mu\alpha_{1}'{(e')^{2}}}{\hbar^{2}r^4}+
\frac{\mu\left(\alpha_{2}'-6a_{0}\beta_{1}'
\right)(e')^2}{\hbar^{2} r^6}\right]\chi=0.
\end{eqnarray}
Therefore, for neutral targets ($Z=z'$) this is a particular
case of the Ince limit of the DCHE, as we see from Eq. (\ref{N2})
with $\omega^2=0$, $2\eta\omega=-q$ and $z=\rho^2=r^2$.
On the hand, if the inverse sixth-power term vanishes
($\alpha_{2}^{'}=6a_{0}\beta_{1}^{'}$),
this radial Schr\"odinger equation is a particular
case of the DCHE as seen from Eq. (\ref{N1}) with $B_{2}=2$,
for neutral or ionized targets.
In both cases the energy of the incident particle ($k^2$) is given
and, consequently, there is no free parameter in these
equations since the other constants are also fixed.
Then, convergent solutions require
a phase parameter  $\nu$, analogously to the scattering by
the field of an electric dipole \cite{leaver1}.
To obtain the radial dependence $R(r)$ we must convert
Eq. (\ref{radial2}) into the DCHE (\ref{dche}) and its
limit (\ref{lindemann2}). Below we discuss only the asymptotic
behaviors of the solutions for the each case. By this reason,
we do not write the recurrence relations for the coefficients.\\

\noindent
{\bf Potential with inverse fourth and sixth-power terms}. Eq. (\ref{N2})
suggests the substitutions
\begin{eqnarray*}
\begin{array}{l}
z= r^2, \ \ \chi(r)=e^{-B_{1}/(2r^2)}r^{B_{2}-(1/2)}U(z=r^2)\ \
\mbox{with}
\vspace{3mm}\\
B_{1}=\pm\frac{e'}{\hbar}\sqrt{\mu(6a_{0}\beta_{1}^{'}-
\alpha_{2}^{'})}, \ \ B_{2}=2-\frac{\alpha_{1}^{'}(e')^2}{2\hbar^2
B_{1}}, \ \ (6a_{0}\beta_{1}^{'}\neq\alpha_{2}^{'})
\end{array}
\end{eqnarray*}
which transform the Schr\"odinger equation (\ref{radial2}) into
\begin{eqnarray*}
z^2\frac{d^{2}U}{dz^{2}}+(B_{1}+B_{2}z)
\frac{dU}{dz}+
\left[\left( \frac{B_{2}}{2}-\frac{1}{4}\right)
\left( \frac{B_{2}}{2}-\frac{3}{4}\right)-\frac{l(l+1)}{4}+
\frac{k^2}{4}z-\frac{\mu}{2}(Z-z')\sqrt{z}\right]U=0.
\end{eqnarray*}
Then, for  $Z\neq z'$, the  Schr\"odinger
equation is more general than the Ince limit of DCHE.
However, assuming a neutral target, we may form two
pairs of solutions according to
\begin{eqnarray}
R_{i\nu}(r)=\frac{1}{r}
\chi_{i\nu}(r) =e^{-B_{1}/(2r^2)}r^{B_{2}-(3/2)}
U_{i\nu}(z=r^2)  \ (i=1,2)
\end{eqnarray}
where on the right-hand side  the $U_{i\nu}$ represent the solutions
with a phase parameter for the Ince limit of the DCHE,
given in section 3.1. Then, taking into account that for this
case $q=k^2/4=\mu E/(2\hbar^2)$ and $z=r^2$, we find
\begin{eqnarray}
\begin{array}{l}
R_{1\nu}^{0}(r)=e^{-B_{1}/(2r^2)}
r^{-2\nu-\frac{3}{2}}\displaystyle \sum_{n=-\infty}^{\infty}
b_{n}^{(1)}
\left(\frac{B_{1}}{r^2}\right)^{n}\Psi\left(n+\nu+
\frac{B_{2}}{2},2n+2\nu+2;
\frac{B_{1}}{r^2}\right),
\vspace{.3cm}\\
R_{1\nu}^{\infty}(r) =e^{-B_{1}/(2r^2)}r^{-1/2}
\displaystyle \sum_{n=-\infty}^{\infty} b_{n}^{(1)}
K_{2n+2\nu+1}\left(\pm ikr\right);
\end{array}
\end{eqnarray}
\begin{eqnarray}
\begin{array}{l}
R_{2\nu}^{0}(r)=e^{{B_{1}}/{(2r^2)}}r^{-2\nu-
\frac{3}{2}}\displaystyle \sum_{n=-\infty}^{\infty}b_{n}^{(2)}
\left(-\frac{B_{1}}{r^2}\right)^{n}\Psi\left(n+\nu+2-
\frac{B_{2}}{2},2n+2\nu+2;
-\frac{B_{1}}{r^2}\right),
\vspace{.3cm}\\
R_{2\nu}^{\infty}(r)=e^{{B_{1}}/{(2r^2)}}r^{-1/2}
\displaystyle \sum_{n=-\infty}^{\infty}b_{n}^{(2)}
K_{2n+2\nu+1}\left(\pm ikr \right).
\end{array}
\end{eqnarray}
From these expressions we obtain
\begin{eqnarray}
\lim_{r\rightarrow\infty}R_{1\nu}^{\infty}(r)\propto
\lim_{r\rightarrow\infty}R_{2\nu}^{\infty}(r)
\sim\frac{e^{\mp ikr}}{r}, \ \ -\frac{3\pi}{2}<\arg{(\pm ikr)}<
\frac{3\pi}{2}
\end{eqnarray}
where we have employed the limit (\ref{21}) for the modified Bessel
functions. Thus, when $r\rightarrow\infty$,
the solutions $R_{i\nu}^{\infty}$ are bounded even if
$k$ is a pure imaginary, since in this case $\exp(ikr)$ or $\exp(-ikr)$
goes to zero. At $r=0$, Eq. (\ref{tricomi}) implies that
\begin{eqnarray}\begin{array}{l}
\displaystyle\lim_{r\rightarrow0}R_{1\nu}^{0}(r)
\sim e^{-B_{1}/r^2}r^{B_{2}-(3/2)}, \ \ -\frac{3\pi}{2}<\arg{\frac{B_{1}}{r^2}}<
\frac{3\pi}{2},
\vspace{3mm}\\
\displaystyle\lim_{r\rightarrow0}R_{2\nu}^{0}(r)
\sim e^{B_{1}/r^2}r^{(5/2)-B_{2}}, \ \ -\frac{3\pi}{2}<\arg{\left( -\frac{B_{1}}{r^2}\right) }<
\frac{3\pi}{2}.
\end{array}
\end{eqnarray}
Thence, if $B_{1}$ is a positive real number, the first limit goes to
zero; if $B_{1}$ is a negative real number, the second limit goes to
zero. However, if $B_{1}$ is a pure imaginary, we write
\begin{eqnarray*}
B_{1}=iC, \ \ B_{2}=2+\frac{i\alpha_{1}^{'}(e')^2}{2\hbar^2C}
\end{eqnarray*}
where  $C $ is real. Thus we find
\begin{eqnarray*}
\vert R_{1\nu}^{0}(r)\vert\propto\vert R_{2\nu}^{0}(r)\vert\sim\sqrt{r}\rightarrow 0.
\end{eqnarray*}
Therefore, it is possible to find at least one pair of solutions
for which both the solutions are bounded at the singularities.\\

%
\noindent
{\bf Potential without inverse sixth-power term}.
From Eq. (\ref{N1}) we find that the substitutions
\begin{eqnarray*}
z= r, \ \ \chi(r)=e^{-B_{1}/(2r)}r^{B_{2}/2}U(z=r)\ \
\mbox{with}\ \
\hbar^2B_{1}^{2}=-4\mu(e')^2, \ \ B_{2}=2
\end{eqnarray*}
transform the Schr\"odinger equation (\ref{radial2}) into
\begin{eqnarray}
r^2\frac{d^{2}U}{dr^{2}}+(B_{1}+2r)
\frac{dU}{dr}+
\left[-l(l+1)-
\frac{2\mu}{\hbar^2}(Z-z')\overline{e}e'r+k^2 r^ 2+
\frac{\mu(\alpha_{2}^{'}-6a_{0}\beta_{1}^{'})(e')^2}
{\hbar^2r^4}\right]U=0.
\end{eqnarray}
Thus, in absence of the inverse sixth-power term,
the radial Schr\"odinger equation, even if we have a Coulomb term
in the potential, may be solved by
\begin{eqnarray}
R_{i\nu}(r)=e^{-B_{1}/(2r)}U_{i\nu}(z=r)
\end{eqnarray}
where  $U_{i\nu}(z=r)$ are solutions with a phase parameter
for the DCHE with $z=r$ and $B_{2}=2$ (see Appendix B). As
\begin{eqnarray}
\omega=\pm k\leftrightarrow
\pm\eta=\pm\frac{\mu}{k\hbar^2}(Z-z')\bar{e}e',\ \ k=\frac{\sqrt{2\mu E}}{\hbar},
\end{eqnarray}
those solutions give
\begin{eqnarray}
\begin{array}{l}
R_{1\nu}^{0} (r) =e^{\pm ikr-\frac{B_{1}}{2r}}\displaystyle \sum_{n=-\infty}^{\infty}b_{n}
\left(\frac{B_{1}}{r}\right)^{n+\nu+1}\Psi\left(n+\nu+1,2n+2\nu+2;
\frac{B_{1}}{r}\right),
\vspace{0.3cm}\\
R_{1\nu}^{\infty}(r) =e^{\pm ikr-\frac{B_{1}}{2r}}
\displaystyle \sum_{n=-\infty}^{\infty}b_{n}(\mp2ikr)^{n+\nu}\Psi(n+\nu+1\pm
i\eta,2n+2\nu+2;\mp 2ik r);
\end{array}
\end{eqnarray}
\begin{eqnarray}
\begin{array}{l}
R_{2\nu}^{0} (r) =e^{\pm ik r+\frac{B_{1}}{2r}}\displaystyle \sum_{n=-\infty}^{\infty}b_{n}
\left(-\frac{B_{1}}{r}\right)^{n+\nu+1}\Psi\left(n+\nu+1,2n+2\nu+2;
-\frac{B_{1}}{r}\right),
\vspace{3mm}\\
R_{2\nu}^{\infty}(r) =e^{\pm ik r+\frac{B_{1}}{2r}}\displaystyle \sum_{n=-\infty}^{\infty}b_{n}
(\mp 2ik r)^{n+\nu}\Psi(n+\nu+1\pm
i\eta,2n+2\nu+2;\mp 2ik r).
\end{array}
\end{eqnarray}
Using Eq. (\ref{tricomi}), we find
\begin{eqnarray}
\lim_{r\rightarrow\infty}R_{1\nu}^{\infty}(r)\propto
\lim_{r\rightarrow\infty}R_{2\nu}^{\infty}(r)
\sim r^{\mp i\eta}\ \frac{e^{\pm ikr}}{r}, \ \ -\frac{3\pi}{2}<\arg{(\mp ikr)}<
\frac{3\pi}{2}
\end{eqnarray}
Thus, when $r\rightarrow\infty$,
the solutions $R_{i\nu}^{\infty}$ are bounded even if
$k$ is a pure imaginary number, since in this case the behavior
of $\exp(ikr)$ or $\exp(-ikr)$
predominates over the other factor. At $r=0$, by using
Eq. (\ref{tricomi}) we get
\begin{eqnarray}\begin{array}{l}
\displaystyle\lim_{r\rightarrow0}R_{1\nu}^{0}(r)
\sim e^{-B_{1}/(2r)}, \ \ -\frac{3\pi}{2}<\arg{\frac{B_{1}}{r}}<
\frac{3\pi}{2}
\vspace{3mm}\\
\displaystyle\lim_{r\rightarrow0}R_{2\nu}^{0}(r)
\sim e^{B_{1}/(2r)}, \ \ -\frac{3\pi}{2}<\arg{\left( -\frac{ B_{1}}{r}\right) }<
\frac{3\pi}{2}.
\end{array}
\end{eqnarray}
As $B_{1}$ is a pure imaginary number, we find that
\begin{eqnarray*}
\vert R_{1\nu}^{0}(r)\vert\propto\vert R_{2\nu}^{0}(r)\vert\sim 1
\end{eqnarray*}
Therefore, in this case we can form two pairs
of solutions which are regular at the singular points, both pairs
having the same series coefficients. For neutral targets
($\eta=0$) the previous results have already been found by B\"uhring
who has treated the Schr\"odinger equation as a
DCHE \cite{buhring2,buhring1}. Before this author, the  Schr\"odinger
equation  (for neutral targets and
an inverse fourth-power polarization potential) had been
transformed into a Mathieu  equation \cite{vogt,holzwarth}.
Thus, the B\"uhring approach is profitable since it works
for ionized targets, too. In addition, as we have seen,
for inverse sixth-power polarization potential, the Schr\"odinger
equation may be transformed  to the Ince limit of the DCHE,
provided that the target is neutral.

\section*{5. Final remarks}
We have constructed the differential equation (\ref{lindemann})
by applying the Ince limit, defined in Eq. (\ref{limits}), to a generalized
spheroidal wave equation (GSWE). The Leaver limit
($z_{0}\rightarrow 0$) of that equation
has afforded Eq. (\ref{lindemann2}) that turns out to be the Ince limit of
a double-confluent Heun equation (DCHE) as well. The subnormal
Thom\'e behavior at $z=\infty$, for the solutions of the
these Ince limits of the GSWE and DCHE, distinguishes such
equations from the original GSWE and DCHE hitherto considered
in the literature.

In section 2, a pair of solutions (with a phase parameter)
for the Ince limit of the GSWE has been found
as the Ince limit of a pair of solutions for
the original GSWE. One solution is given by a series of
hypergeometric functions and the other by
a series of modified Bessel functions of the second
kind. Both solutions in that pair have
the same series coefficients but different regions of convergence,
as in solutions for the Mathieu
equations. Other pair has followed from the first one by means of
a transformation rule. Hence, four pairs of solutions without phase
parameter have resulted from the truncation of the
series with a phase parameter, that is, by restricting the summation
index of the series to $n\geq0$.

In section 3, solutions for the Ince limit of the DCHE have
been established by taking the Leaver limit
of solutions for the Ince limit of the GSWE. These solutions,
given by series of irregular confluent hypergeometric functions
and modified Bessel functions,  present the
appropriate behavior at the irregular singularities
$z=0$ and $z=\infty$. Note, nonetheless, that  in sections 3 and 4
we have dealt with expansions in
series of modified Bessel functions only.
Other possibilities may be investigated,
specially solutions in series of Bessel function products,
as these could have important properties as regards
the convergence of the series.

In the solutions without phase parameter for the Ince limits
of the GSWE and DCHE, there are  three possible forms to the
recurrence relations for the series coefficients. This fact is
relevant in itself and, in particular, is essential to
recover solutions for the Mathieu equation from the ones for the
Ince limit of the GSWE.

The solutions we have obtained for the Mathieu equation
are already known and exhibit the usual parity
and periodicity properties. This includes also
the solutions found by Poole, given by two-sided series
($-\infty<n<\infty$) and having period $2\pi m$, where $m$ is any integer
greater than $1$. However, we note that other types of
solutions for the Mathieu equations (and also for the
Whittaker-Hill equations)
are possible, since these equations may be considered
as particular cases of both the GSWE and
double-confluent Heun equations as well \cite{decarreau2}.

At last, notice that we have point out no
application for Ince limit of the GSWE. Nevertheless, in section
4 we have seen that
the Schr\"odinger equation (\ref{radial2}) for the scattering
of low-energy particles by polarizable targets leads to an
DCHE and its Ince limit.
The exception is the Schr\"odinger equation  with
Coulomb and inverse sixth-power terms which requires solutions
for a more general equation, possibly similar to
an equation considered by Kurth and Schmidt in \cite{kurth}.

I thank Herman J. Mosquera Cuesta for his careful reading
of this manuscript and valuable suggestions. I also thank the
participants of the ICRA-BR ``Pequenos Semin\'arios" for discussions
and insight on potential physical applications of the results of the
present investigation.
%
\section*{Appendix A: Degenerate DCHEs}
\protect\label{A}
\setcounter{equation}{0}
\renewcommand{\theequation}{A\arabic{equation}}
Let us show that DCHE
\begin{eqnarray*}
z^2\frac{d^{2}U}{dz^{2}}+(B_{1}+B_{2}z)\frac{dU}{dz}+
\left(B_{3}-2\eta\omega z+\omega^{2}z^2\right)U=0,\ \ (B_{1}\neq 0,\
\ \omega\neq 0),
\end{eqnarray*}
for $B_{1}=0$
and/or $\omega=0$ degenerates into a confluent
hypergeometric equation or an equation with
constant coefficients. Thus, if $B_{1}=0$
and $\omega\neq 0$, the
substitutions
\begin{eqnarray*}
 y=-2i\omega z, \ \  U(z)=e^{-y/2}y^{\alpha}f(y), \ \
\alpha^2-(1-B_{2})\alpha+B_{3}=0
\end{eqnarray*}
give the confluent hypergeometric equation
\begin{eqnarray*}
y\frac{d^2f}{dy^2}+[(2\alpha+B_{2})-y]\frac{df}{dy}-
\left(i\eta+\alpha+\frac{B_{2}}{2}\right)f=0.
\end{eqnarray*}
If $B_{1}\neq 0$ and $\omega= 0$, the change of variables
\begin{eqnarray*}
y=B_{1}/z, \ \  U(z)=y^{\beta}g(y), \ \ \beta^2-(B_{2}-1)\beta+B_{3}=0
\end{eqnarray*}
leads to
\begin{eqnarray*}
&&y\frac{d^2g}{dy^2}+[(2\beta+2-B_{2})-y]\frac{dg}{dy}-\beta g=0.
\end{eqnarray*}
If $B_{1}=\omega=0$, we find an equation with constant coefficients by
taking $z=\exp{y}$.

Now let us show that the Ince limit of the DCHE
\begin{eqnarray*}
z^2\frac{d^{2}U}{dz^{2}}+(B_{1}+B_{2}z)\frac{dU}{dz}+
\left(B_{3}+q z\right)U=0,\ (q\neq0,\ B_{1}\neq 0)
\end{eqnarray*}
also gives degenerate cases if $q\neq0$ and/or $\ B_{1}\neq 0$.
In fact, if $q=0$ and $B_{1}\neq 0$, this equation is equivalent
to the DCHE with $\omega= 0$ and $B_{1}\neq 0$.
If $q\neq0$ and $B_{1}= 0$, the
substitutions
\begin{eqnarray*}
\xi=\pm 2i\sqrt{qz}
,\ \ U(z)=\xi^{1-B_{2}}T(\xi)
\end{eqnarray*}
reduces the equation to the modified Bessel equation
\begin{eqnarray*}
\xi^2\frac{d^2T}{d\xi^2}+\xi\frac{dT}{d\xi}-
\left[(1-B_{2})^2-4B_{3}+\xi^2\right]T=0.
\end{eqnarray*}
Finally, for $q=B_{1}= 0$, we find again an equation with constant
coefficients by taking $z=\exp{y}$.
%
%
\section*{Appendix B: The solutions in series of Bessel functions}
\protect\label{B}
\setcounter{equation}{0}
\renewcommand{\theequation}{B\arabic{equation}}
The solution $U_{1\nu}^{\infty}(z)$ in
series of Bessel functions can also be constructed as follows.
We perform the substitutions
\begin{eqnarray}\label{2.2}
\xi=\pm2i\sqrt{qz},\ \ U(z)=\xi^{1-B_{2}}Y(\xi)
\end{eqnarray}
in the Ince limit of the GSWE (\ref{lindemann}). This yields
\begin{eqnarray}\label{2.4}
\xi^2\frac{d^2Y}{d\xi^2}+\xi\frac{dY}{d\xi}-
\xi^2Y=-4qz_{0}\frac{d^2Y}{d\xi^2}-
\frac{4q(z_{0}-2B_{1}-2B_{2}z_{0})}{\xi}
\frac{dY}{d\xi} \nonumber\\
+\left[4q(1-B_{2})\frac{2B_{1}+B_{2}z_{0}+z_{0}}
{\xi^2}+(1-B_{2})^2+4qz_{0}-4B_{3}\right]Y.
\end{eqnarray}
Now we expand $Y(\xi)$ according to
\begin{eqnarray}\label{2.6}
Y(\xi)=\displaystyle \sum_{n=-\infty}^{\infty}b_{n}^{(1)}
K_{\lambda}(\xi), \ \ \lambda=2n+2\nu+1,
\end{eqnarray}
where $K_{\lambda}(\xi)$ denotes the
modified Bessel function of the second
kind \cite{luke}. The last equation and (\ref{2.2})
afford the solution $U_{1\nu}^{\infty}(z)$.

When we insert (\ref{2.6}) into (\ref{2.4}),
we use some difference-differential relations derived
from the properties of $K_{\lambda}$ \cite{luke}.
Thus, we have
\begin{eqnarray*}
\xi^2\frac{d^2K_{\lambda}(\xi)}{d\xi^2}+\xi
\frac{dK_{\lambda}(\xi)}{d\xi}-
\xi^2K_{\lambda}(\xi)=\lambda^2K_{\lambda}(\xi)
\end{eqnarray*}
on the left-hand side and
\begin{eqnarray*}
4\frac{d^2K_{\lambda}(\xi)}{d\xi^2}=
K_{\lambda+2}(\xi)+2K_{\lambda}(\xi)+
K_{\lambda-2}(\xi),\ \
\frac{4}{\xi}\frac{dK_{\lambda}(\xi)}{d\xi}=
-\frac{4\lambda}{\xi^2}K_{\lambda}(\xi)+
\frac{2}{\lambda-1}\left[K_{\lambda-2}(\xi)-
K_{\lambda}(\xi)\right]
\end{eqnarray*}
on the right-hand side. This gives
\begin{eqnarray*}
&&qz_{0}\displaystyle \sum_{n=-\infty}^{\infty}
\left[1+\frac{2[1-2B_{2}-(2B_{1}/z_{0})]}{\lambda-1}\right]b_{n}^{(1)}
K_{\lambda-2}(\xi)\nonumber\\
&&+\displaystyle \sum_{n=-\infty}^{\infty}
\left[\lambda^2+4B_{3}-2qz_{0}-(1-B_{2})^2-
\frac{2qz_{0}[1-2B_{2}-(2B_{1}/z_{0})]}{\lambda-1}
\right]b_{n}^{(1)}
K_{\lambda}(\xi)
\nonumber\\
&&+qz_{0}\displaystyle \sum_{n=-\infty}^{\infty}
b_{n}^{(1)}K_{\lambda+2}(\xi)\nonumber\\
&&=qz_{0}\displaystyle \sum_{n=-\infty}^{\infty}
\left[\left(1-2B_{2}-\frac{2B_{1}}{z_{0}}
\right)\lambda+(1-B_{2})\left(
1+B_{2}+\frac{2B_{1}}{z_{0}}\right)\right]b_{n}^{(1)}
\frac{4K_{\lambda}(\xi)}{\xi^2}.
\end{eqnarray*}
To remove the term $4K_{\lambda}(\xi)/\xi^2$
on the right-hand side we use the relation
\begin{eqnarray*}
\frac{4K_{\lambda}(\xi)}{\xi^2}=
\frac{K_{\lambda-2}(\xi)}{\lambda(\lambda-1)}-
\frac{2K_{\lambda}(\xi)}{(\lambda-1)(\lambda+1)}+
\frac{K_{\lambda+2}(\xi)}{\lambda(\lambda+1)}.
\end{eqnarray*}
Then, reminding that $\lambda=2n+2\nu+1$, we
find
\begin{eqnarray}\label{2.10}
\displaystyle \sum_{n=-\infty}^{\infty}
\alpha_{n-1}^{(1)}b_{n}^{(1)}
K_{2n+2\nu-1}(\xi)+
\displaystyle \sum_{n=-\infty}^{\infty}
\beta_{n}^{(1)}b_{n}^{(1)}
K_{2n+2\nu+1}(\xi)+
\displaystyle \sum_{n=-\infty}^{\infty}
\gamma_{n+1}^{(1)}b_{n}^{(1)}
K_{2n+2\nu+3}(\xi)=0,
\end{eqnarray}
where the coefficients
$\alpha_{n}^{(1)}$, $\beta_{n}^{(1)}$
and $\gamma_{n}^{(1)}$ are
just the ones given in equations (\ref{apB}).
To get the recurrence relations with the form
given in (\ref{rec}), we change $n\rightarrow m+1$ and
$n\rightarrow m-1$ in the first and third
terms, respectively. After this, we equate to zero
the coefficients of each independent
$K_{2m+2\nu+1}(\xi)$.

%
%
On the other hand, to study the convergence of the series, we
apply a Perron-Kreuser theorem \cite{gautschi}
for the minimal solutions of the recurrence relations for
$b_{n}^{(1)}$ and obtain (if $z_{0}\neq0$)
\begin{eqnarray}\label{2.12}
\lim_{n\rightarrow \infty}\frac{b_{n+1}^{(1)}}{b_{n}^{(1)}}=
\lim_{n\rightarrow -\infty}\frac{b_{n-1}^{(1)}}{b_{n}^{(1)}}=
-\frac{qz_{0}}
{4n^2}.
\end{eqnarray}
Using also the relation \cite{luke}
\begin{eqnarray*}
\lim_{\lambda\rightarrow \infty}K_{\lambda}({\xi})=
\frac{1}{2}\Gamma({\lambda})
\left(\frac{\xi}{2}\right)^{-\lambda}
\end{eqnarray*}
and $K_{-\lambda}({\xi})=K_{\lambda}({\xi})$, we get
\begin{eqnarray*}
\lim_{n\rightarrow \infty}
\frac{K_{2n+2\nu+3}({\xi})}{K_{2n+2\nu+1}({\xi})}=
\lim_{n\rightarrow -\infty}
\frac{K_{2n+2\nu-1}({\xi})}{K_{2n+2\nu+1}({\xi})}=
-\frac{4n^2}{qz}.
\end{eqnarray*}
Hence, we have
\begin{eqnarray*}
\lim_{n\rightarrow \infty}\frac{b_{n+1}^{(1)}K_{2n+2\nu+3}({\xi})}
{b_{n}^{(1)}K_{2n+2\nu+1}({\xi})}=
\lim_{n\rightarrow -\infty}\frac{b_{n-1}^{(1)}K_{2n+2\nu-1}({\xi})}
{b_{n}^{(1)}K_{2n+2\nu+1}({\xi})}=
\frac{z_{0}}{z}.
\end{eqnarray*}
Therefore, by the ratio test the series converges for
$|z|>|z_{0}|$. In (\ref{2.12}) we have supposed
that $z_{0}\neq0$ but, if $z_{0}=0$, we find
\begin{eqnarray*}
&&\lim_{n\rightarrow \infty}\frac{b_{n+1}^{(1)}}{b_{n}^{(1)}}=
\lim_{n\rightarrow -\infty}\frac{b_{n-1}^{(1)}}{b_{n}^{(1)}}=
-\frac{B_{1}}
{4n^3}\ \Rightarrow\\
&&
\lim_{n\rightarrow \infty}\frac{b_{n+1}^{(1)}K_{2n+2\nu+3}({\xi})}
{b_{n}^{(1)}K_{2n+2\nu+1}({\xi})}=
\lim_{n\rightarrow -\infty}\frac{b_{n-1}^{(1)}K_{2n+2\nu-1}({\xi})}
{b_{n}^{(1)}K_{2n+2\nu+1}({\xi})}=
\frac{B_{1}}{nz}.
\end{eqnarray*}
Thus, in this limit the series converges for $|z|>0$
and per se this result is already included in
$\vert z\vert>\vert z_{0}\vert$.

%
\section*{Appendix C: Solutions for the DCHE of section 4}
\protect\label{C}
\setcounter{equation}{0}
\renewcommand{\theequation}{C\arabic{equation}}
The Leaver-type solutions for the DCHE (\ref{dche}) present
some simplifications for  $B_{2}=2$.
The solutions given in Ref. \cite{eu} are expansions
in series of regular and irregular confluent hypergeometric functions.
However, to obtain the expected behavior at the singular points
$z=0$ and $z=\infty$, we have to choose the irregular functions.
Then, by using the same notation of sections 2.1 and 3.1,  we find that
for $B_{2}=2$ the first pair of solutions with a phase parameter
is given by
\begin{eqnarray}
\begin{array}{l}
U_{1\nu}^{0} (z) =e^{i\omega z}\displaystyle \sum_{n=-\infty}^{\infty}b_{n}
\left(\frac{B_{1}}{z}\right)^{n+\nu+1}\Psi\left(n+\nu+1,2n+2\nu+2;
\frac{B_{1}}{z}\right),
\vspace{0.3cm}\\
U_{1\nu}^{\infty}(z) =e^{i\omega z}
\displaystyle \sum_{n=-\infty}^{\infty}b_{n}(-2i\omega z)^{n+\nu}\Psi(n+\nu+1+
i\eta,2n+2\nu+2;-2i\omega z),
\end{array}
\end{eqnarray}
and the second pair takes the form
\begin{eqnarray}
\begin{array}{l}
U_{2\nu}^{0} (z) =e^{i\omega z+\frac{B_{1}}{z}}\displaystyle \sum_{n=-\infty}^{\infty}b_{n}
\left(-\frac{B_{1}}{z}\right)^{n+\nu+1}\Psi\left(n+\nu+1,2n+2\nu+2;
-\frac{B_{1}}{z}\right),
\vspace{3mm}\\
U_{2\nu}^{\infty}(z) =e^{i\omega z+\frac{B_{1}}{z}}\displaystyle \sum_{n=-\infty}^{\infty}b_{n}
(-2i\omega z)^{n+\nu}\Psi(n+\nu+1+
i\eta,2n+2\nu+2;-2i\omega z).
\end{array}
\end{eqnarray}
Then, we see that the two pairs have the same
series coefficients $b_{n}$ and the coefficients in the recurrence
relations (\ref{rec}) are simply
\begin{eqnarray}
\begin{array}{l}
\alpha_{n}  =  i\omega B_{1}\left( \frac{n+\nu+1-i\eta}
{2n+2\nu+3}\right) ,\ \
%
\beta_{n}  = B_{3}+(n+\nu)(n+\nu+1),
\gamma_{n}=i\omega B_{1} \left( \frac{n+\nu+
i\eta}{2n+2\nu-1}\right) .
\end{array}
\end{eqnarray}
In these solutions $\nu$ cannot be integer or half-integer and
the $U_{i\nu}^{0}$ converge for any finite $z $,
whereas  the $U_{i\nu}^{\infty}$ converge for $\vert z\vert>0$.
Note, moreover,  that the irregular confluent hypergeometric functions
that appear in $U_{i\nu}^{0}$ could be rewritten in terms of
modified Bessel of the second kind by using the definition (\ref{2.8}).
In the solutions $U_{i\nu}^{\infty}$ the confluent hypergeometric
functions could be rewritten in terms of the Hankel functions
$H_{\rho}^{(1)}$ but only if $\eta=0$ (neutral target, in
the problem of section 4). For this we have to use the relation
\cite{erdelyi1}
\begin{eqnarray*}
\Psi\left(\rho+\frac{1}{2},2\rho+1;-2ix\right)=
\frac{i}{2\sqrt{\pi}}
e^{i(\rho \pi-x)}H_{\rho}^{(1)}(x), \ \rho=n+\nu+\frac{1}{2}.
\end{eqnarray*}
The asymptotic behaviors of the solutions given in (C1-2) may
be found by using Eq. (\ref{tricomi}).
%
%
%

\end{document}